\documentclass[twocolumn,english,prl]{revtex4-1}
\usepackage{color}
\usepackage{graphicx}
\usepackage[T1]{fontenc}
\usepackage{float}
\usepackage{textcomp}
\usepackage{amsmath}
\usepackage{amssymb}
\usepackage{graphicx}
\usepackage{esint}
\usepackage{natbib}
\usepackage{color}

\begin{document}

\title{Electrical control of a confined electron spin in a silicene quantum dot}


\author{Bart\l{}omiej Szafran}

\affiliation{AGH University of Science and Technology, Faculty of Physics and
Applied Computer Science,\\
 al. Mickiewicza 30, \\30-059 Krak\'ow, Poland}

\author{Alina Mre\'{n}ca-Kolasi\'{n}ska }

\affiliation{AGH University of Science and Technology, Faculty of Physics and
Applied Computer Science,\\
 al. Mickiewicza 30, \\30-059 Krak\'ow, Poland}

\author{Bart\l{}omiej Rzeszotarski}

\affiliation{AGH University of Science and Technology, Faculty of Physics and
Applied Computer Science,\\
 al. Mickiewicza 30, \\30-059 Krak\'ow, Poland}

\author{Dariusz \.Zebrowski}

\affiliation{AGH University of Science and Technology, Faculty of Physics and
Applied Computer Science,\\
 al. Mickiewicza 30, \\30-059 Krak\'ow, Poland}

\begin{abstract}
We study spin control for an electron confined 
in a flake of silicene. 
We find that the lowest-energy conduction-band levels are split by the diagonal intrinsic spin-orbit coupling
into Kramers doublets with a definite projection of the spin on the orbital magnetic moment. 
We study the  spin control by AC electric fields using the non-diagonal Rashba component
of the spin-orbit interactions with the time-dependent atomistic tight-binding approach. The Rashba interactions 
in AC electric fields produce Rabi spin-flips times of the order of a nanosecond.
These  times can be reduced to tens of picoseconds
provided that the vertical electric field is tuned to an avoided crossing open by the Rashba spin-orbit interaction. 
We demonstrate that the speedup of the spin transitions is possible due to the intervalley coupling induced by the armchair edge of the flake.
The study is confronted with the results for circular quantum dots decoupled from the edge with well defined angular momentum and valley index.
\end{abstract}

\maketitle

\section{Introduction}

Silicene \cite{chow} is potentially an attractive alternative of graphene \cite{Neto09} for spintronic \cite{Zutic04} applications.
 The material is characterized by strong intrinsic spin-orbit coupling \cite{Liu11,Liu,Ezawa} that should allow for observation
of quantum spin Hall effect \cite{Liu11}.
Moreover, anomalous Hall effect \cite{Ezawa} and its valley-polarized variant \cite{Pan14} as well as giant magnetoresistance \cite{Xu12,Rachel14}
were predicted for silicene systems. Possible applications to spin-filtering \cite{Tsai13,szak15,miso15,nunez16,wu15} were proposed
and topological phase transitions in the edge states driven by the perpendicular electric field are expected \cite{Ezawa12a, Tabert13,romera}.
An operating room-temperature field effect transistor has recently been demonstrated \cite{Tao15} for
 silicene 
transfered on Al$_2$O$_3$ dielectric,
which relatively weakly perturbs the band structure of silicene near the Dirac points \cite{al2o3}.
Deposition  on a non-metallic surface \cite{Tao15,al2o3,nonmetal,nonmetal1,nonmetal2,nonmetal3} is necessary, since the metal substrate \cite{Vogt12,Aufray10,Feng12,me1,me2,me3} masks the electronic properties of silicene.

Spin-orbit coupling is relevant for manipulation and control 
of single carrier spins confined in quantum dots for applications in quantum information processing \cite{Zutic04,Komputer,dl,kl}
 since it allows for addressing individual  quantum dots by electric fields. 
In particular,  control of the confined spin  by AC electric fields is possible with the Rashba spin-orbit coupling that translates the electron motion in space into an effective magnetic field \cite{meier} that drives the spin rotations  \cite{edsr1,edsr2,edsr3}.  
This procedure, known as the electric-dipole spin resonance (EDSR)   \cite{edsr1,edsr2,edsr3},
 was implemented in III-V semiconductors \cite{edsr4,edsr5,edsr6,edsr7,extreme,stroer} and in carbon nanotubes \cite{edsrcnt1,edsrcnt2,edsrcnt3} for studies of the spin-related properties 
of confined electron systems. 
 In the present paper we study the spin-orbit coupling in a silicene flake as a resource for  EDSR
and we simulate the confined spin flips driven by oscillating electric fields.

The carrier eigenstates in graphene flakes have been extensively studied in the literature \cite{gru,zarenia,flake1,flake2,flake3,flake4,flake5,mori}.
Besides the significant spin-orbit coupling \cite{Liu,Ezawa} silicene 
 due to the buckling of the crystal lattice \cite{buck0,Cinquanta12,chow} allows for  control of the electronic structure with perpendicular electric fields.
 The perpendicular electric field  opens the energy gap \cite{ni,Drummond12}  and shifts the energies of the edge-localized states  \cite{abdelsalam,kiku}.
The electric fields of the order of 1V/\AA\; are necessary
for the silicene bandgap tuning \cite{Drummond12} in particular for applications to room temperature field effect transistors \cite{ni}
or to spin-filtering \cite{Tsai13}.

For the purpose of the present study we employ a hexagonal flake with armchair boundaries. The armchair termination does not support edge states \cite{zarenia}
and the energy spectrum  contains a well resolved energy gap due to the quantum confinement. The energy gap allows us to focus on a single excess electron 
 confined within the flake
and adopt  a frozen-valence-band approximation when  AC electric fields are applied for EDSR. 
We find that the spin-orbit coupling component which governs the spectrum is the intrinsic  contribution of Kane-Mele \cite{km} form which is spin-diagonal
and splits the fourfold degenerate ground-state into Kramers doublets with definite projections of the spin on the orbital magnetic moment.
In presence of strong perpendicular electric fields 
 the typical spin-flip transitions  times are of the order of a nanosecond, and the transitions occur according to the two-level Rabi resonance mechanism.
We show that the spin transition times  are nonlinear and nonmonotonic functions of the vertical electric field and can be significantly reduced within avoided crossings opened by the Rashba interaction. 
These avoided crossing occurs in the low-energy spectra in presence of intervalley coupling that is introduced by the armchair edge of the flake \cite{revrev}. The conclusion is drawn from modeling of circular quantum dots with a well defined valley index and angular momentum with both - the atomistic tight-binding and the Dirac approximation to the Hamiltonian.
Formation of spin-valley doublets by the intrinsic spin-orbit interaction for lifted valley scattering is also discussed.

\section{Theory}
The band structure for silicene deposited on Al$_2$O$_3$ \cite{Tao15}  is close to the one of the free-standing silicene \cite{al2o3} near the charge neutrality point.
For that reason we use the atomistic tight-binding Hamiltonian  for free-standing silicene  \cite{Liu} in the basis spanned by the $p_z$ spin-orbitals, 
\begin{eqnarray}
H_0&=&-t\sum_{\langle k,l\rangle \alpha }  c_{k\alpha}^\dagger c_{l\alpha} +it_2 \sum_{\langle \langle k,l\rangle \rangle \alpha, \beta } \nu_{kl} c^\dagger_{k\alpha} \sigma^{z}_{\alpha\beta}c_{l\beta} \nonumber \\&& -it_1 \sum_{\langle \langle  k,l \rangle \rangle \alpha,\beta } \mu_{kl} c^\dagger _{k\alpha}\left(\vec{\sigma}\times\vec{d}_{kl} \right)^z_{\alpha\beta} c_{l\beta} \nonumber 
\\ && +it_3 (F_z) \sum_{\langle k,l \rangle,\alpha,\beta}  c_{k\alpha}^\dagger \left(\vec{\sigma}\times\vec{d}_{kl} \right)^z_{\alpha\beta}  c_{l\beta}  \nonumber 
\\ && +e F_z \sum_{k,\alpha} z_k c^\dagger_{k\alpha}c_{k\alpha},  \label{hb0}
\end{eqnarray}
where the indices  $k,l$ run over the ions while $\alpha$ and $\beta$  over the spin degree of freedom. In Eq. (\ref{hb0}) $\langle k,l\rangle $ stands for summation over the nearest neighbor ions and $\langle\langle k,l\rangle\rangle $ for the next nearest neighbors. The first term of the Hamiltonian describes the nearest neighbor hopping with $t=1.6$ eV \cite{Liu,Ezawa}. The second term with $t_2$ 
is the intrinsic spin-orbit interaction \cite{km,km2} with $\nu_{kl}=+1$  ($\nu_{kl}=-1$) for the counterclockwise (clockwise) next-nearest neighbor hopping. The adopted value \cite{Liu,Ezawa} of the intrinsic spin-orbit parameter is $t_2=3.9/3\sqrt{3}$ meV. 
The third term in Eq. (\ref{hb0}) with $t_1$ introduces the  Rashba \cite{Liu} interaction due the built-in electric field
which results from the vertical shift of the A and B sublattices in silicene, with ${\bf d}_{kl}=\frac{{\bf r}_l-{\bf r_k}}{|{\bf r}_l-{\bf r_k}|}$ 
and ${\bf r_k}=(x_k,y_k,z_k)$ the position of the $k$-th ion.
The intrinsic Rashba interaction acts for the next nearest neighbors $\langle \langle k,l\rangle \rangle$   with $\mu_{kl}=+1$ for the  sublattice A and $\mu_{kl}=-1$ for the 
sublattice B.
For the intrinsic Rashba parameter we take $t_1=\frac{2\times 0.7}{3} $ meV \cite{Liu,Ezawa}. The Hamiltonian component with $t_3$ is the extrinsic
Rashba term which results from the external electric field perpendicular to the silicene plane or the  mirror symmetry broken by e.g. the substrate.
The parameter $t_3$ varies linearly with the external field
with $t_3(F_z)=10$ $\mu$eV for $F_z=17$ meV/\AA\; \cite{Ezawa}. The last term of Hamiltonian (\ref{hb0}) introduces the electrostatic potential due to the perpendicular electric field with $z_k=\pm \frac{1}{2}l $ with plus for ion $k$ in the $A$ sublattice and minus for the $B$ sublattice, and $l=0.46$\; \AA \;
is the vertical shift of the $A$ and $B$ sublattice planes.  

We show below that in order to activate effective spin transitions the perpendicular electric
field of the order of 1 V/\AA\; is required. A field this high  can be obtained for silicene sandwiched within dielectric 
between  metal gates \cite{ni}.  The external Rashba interaction introduces
 also the effect of the substrate. In particular, the energy gap found experimentally in Ref. \cite{Tao15} of 210 meV corresponds to an effective vertical electric field of the order of $\simeq$ 1.2 V/\AA \;
according to a linear extrapolation of the data of Ref. \onlinecite{ni}.


\begin{figure}
(a) \includegraphics[width=.6\columnwidth]{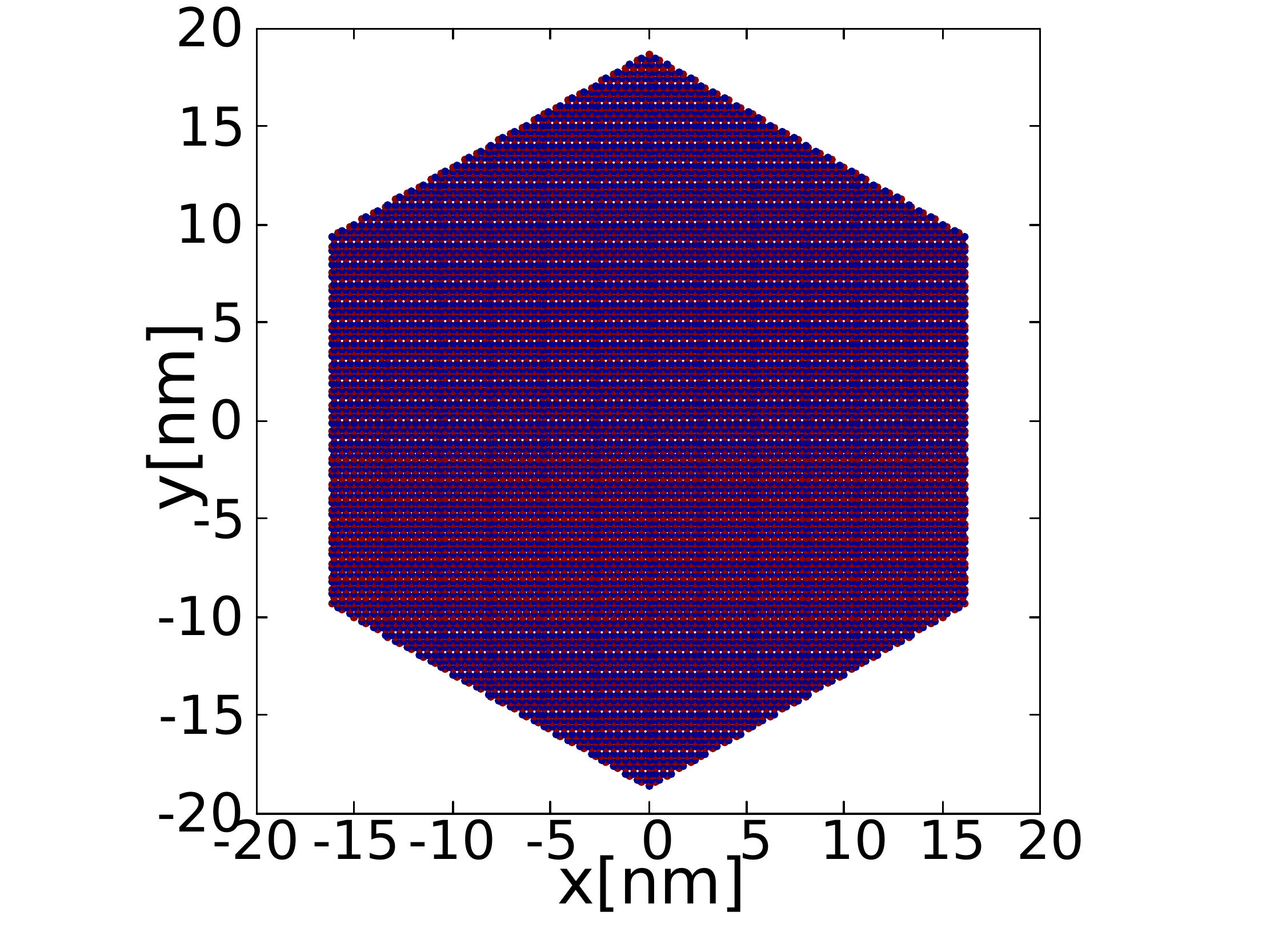} \\
(b) \includegraphics[width=.6\columnwidth]{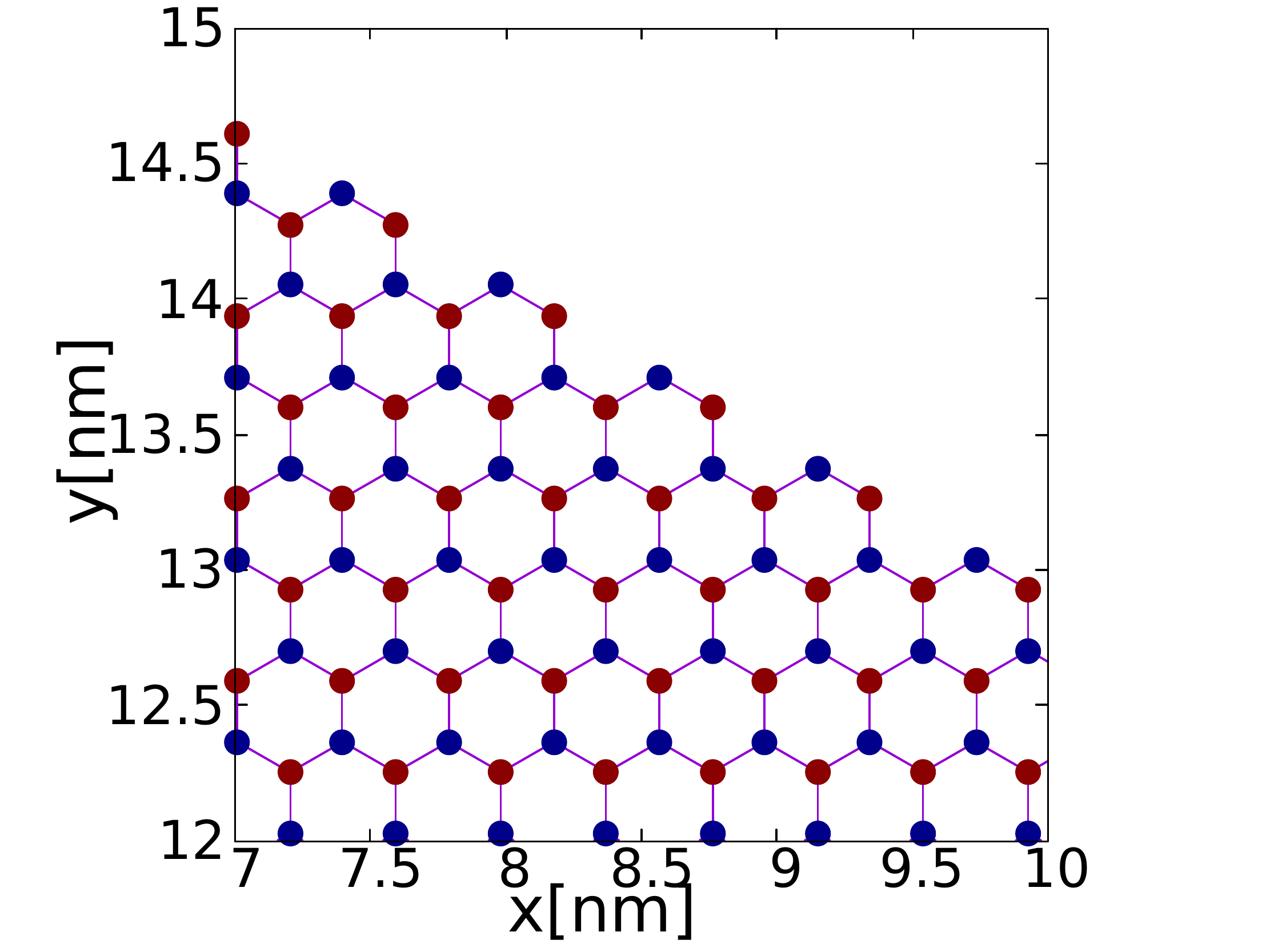}  
\caption{(a) The hexagonal silicene flake considered in this work. The side length of the flake is 18.64 nm, unless stated otherwise. 
(b) Zoom at the edge of the flake. Blue and red dots indicate the A and B sublattices. The sublattices are displaced by $l=0.46$ \AA\; in the $z$ direction. The lattice constant with $a=3.893$\; \AA \;is adopted.
} \label{sche}
\end{figure}

\begin{figure}
\includegraphics[width=0.9\columnwidth]{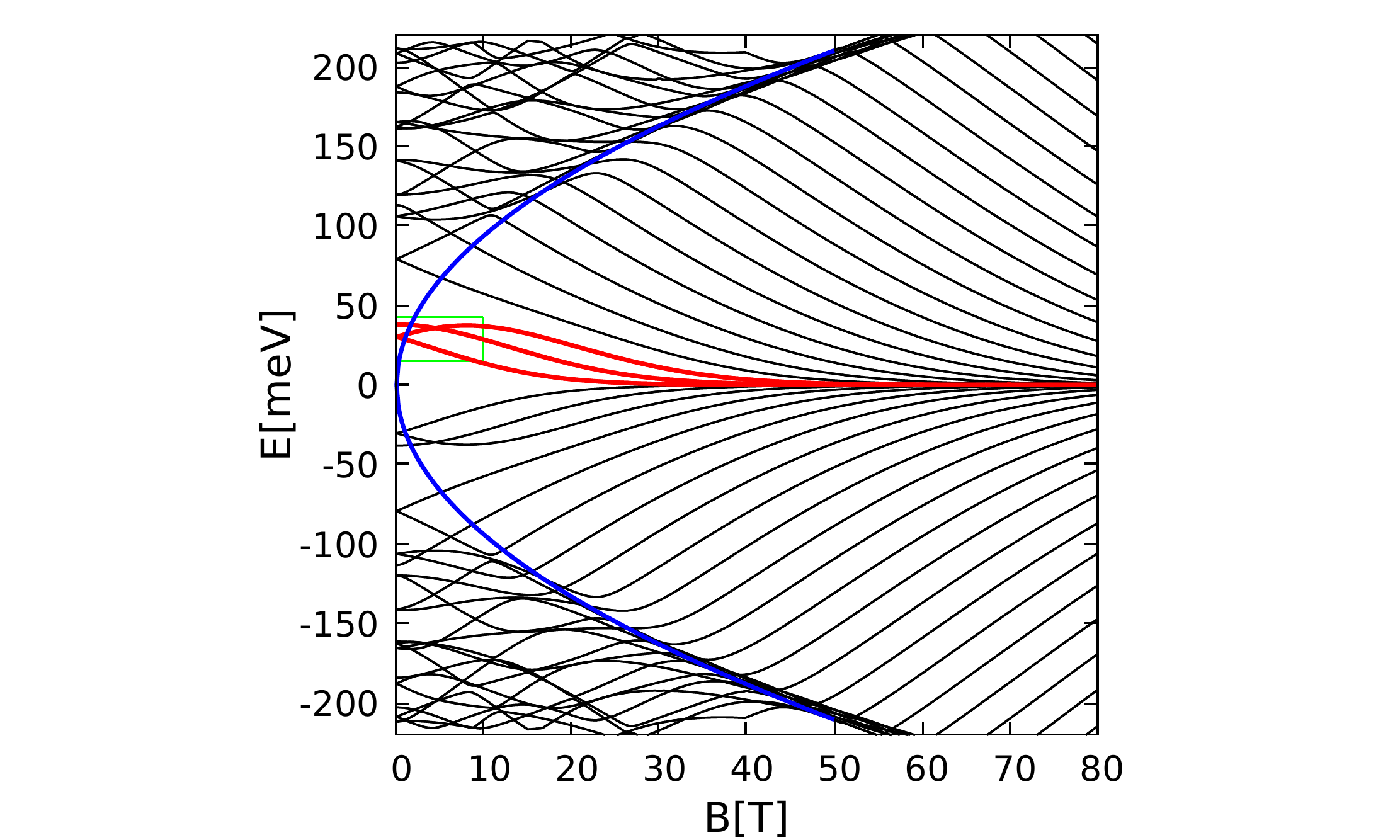} 
\caption{The energy spectrum for the silicene flake as a function of the perpendicular magnetic field
in the absence of the external electric field and spin-orbit interaction. Here and only here 
the Zeeman interaction is neglected.
With the red lines we plotted the lowest-energy states of the conduction band that we consider in detail in Section III.
In Section III, we focus on the magnetic field and energy range that is marked with the green rectangle.
The blue lines indicate the Landau levels with $l=\pm 1$: $E_{LL}=\pm \frac{3dt}{2l_B}\sqrt {2|l|}$, with the magnetic length $l_B=\sqrt{\frac{\hbar}{eB}}$, and the
nearest Si-Si in-plane distance $d=2.25$\AA.
Zero energy corresponds to both the charge neutrality point and the 0th Landau level. 
} \label{wincy}
\end{figure}

For the Hamiltonian of a general form
\begin{eqnarray} 
H_0=\sum_{k,l,\alpha,\beta} h_{k,\alpha,l,\beta}c_{k\alpha}^\dagger c_{l\beta},   \label{haha}
\end{eqnarray}
with the specific hopping parameters $h_{k,\alpha,l,\beta}$  defined by Eq. (\ref{hb0}), the energy
operator which accounts for the external magnetic field oriented perpendicular to the silicene plane $\vec{B}=(0,0,B)$
follows 
\begin{eqnarray} 
H_B&=&\sum_{k,l,\alpha,\beta} h_{k,l,\alpha,\beta} e^{i\frac{e}{\hbar}\int_{\vec{r_k}}^{\vec{r_l}}\vec {A}\cdot \vec {dl}}c_{k\alpha}^\dagger c_{l\beta} 
\nonumber \\ 
&& +\frac{1}{2}g\mu_B B\sum_{k,\alpha}  \sigma^z_{\alpha,\alpha} c_{k\alpha}^\dagger c_{k\alpha},
\end{eqnarray}
where the first term introduces the Peierls phase, with the vector potential $\vec{A}$ and the second term is the spin Zeeman interaction with the Bohr magneton $\mu_B$ and the electron spin factor $g=2$.

We study the spin flips of the electron confined within the flake by application of an
external in-plane  AC electric field with the time-dependent Hamiltonian 
\begin{equation} H_t=H_B+H'=H_B+eF_{ac}\sum_{k,\alpha}x_k\sin(2\pi\nu t) c_{k\alpha}^\dagger c_{k\alpha},
\end{equation}
where $F_{ac}$ and $\nu$ are the amplitude and the frequency of the AC electric field, respectively. The maximal considered AC amplitude is $F_{ac}=400$ V/cm$ = 4\mu$eV/\AA.

Once the eigenstates $\Psi_n$ of the stationary Hamiltonian are determined $H_B\Psi_n=E_n\Psi_n$, we use
them as the basis for description of the time evolution of the wave function 
\begin{equation} \Psi=\sum_n c_n(t) \exp(-\frac{iE_n t}{\hbar})\Psi_n \label{basis} \end{equation} which when plugged into the Schr\"odinger equation $i\hbar \frac{\partial \Psi}{\partial t}=H_t \Psi$, produces the linear system of differential equations \cite{eosika},
\begin{equation}
i\hbar \frac{dc_k(t)}{dt}=\sum_n c_n(t) eF_{ac} \sin(2\pi\nu t) \langle\Psi_k | x|\Psi_n\rangle e^{-i\frac{E_n-E_k}{\hbar}t}, \label{time}
\end{equation}
that we solve with the implicit trapezoid rule. 

For the purpose of the present study we consider a hexagonal silicene flake given in Fig. \ref{sche}(a) 
 with side length of 18.64 nm.
In the rest of the paper we consider a single excess electron in the conduction band and focus on the low-energy part of the spectrum,
 namely on the energy levels plotted in red within the range indicated in Fig. \ref{wincy} by the green rectangle.
In Figure \ref{wincy}  the spin-orbit coupling was neglected. 
Moreover, Figure \ref{wincy} is the only plot where we neglect the Zeeman interaction.
We note the following: {\it (i)} Without the Zeeman interaction
all levels are degenerate with respect to the spin. {\it (ii)} All the energy levels marked in red tend to 0th Landau level at high $B$.
For the armchair termination of the flake [Fig. \ref{sche}(b)] no localization of the states near the edge is observed \cite{zarenia}
hence the missing zero energy level in Fig. \ref{wincy}. {\it (iii)} A well defined band gap between the conduction and valence band states  near $E=0$.
The maximal potential energy variation  due to the $F_{ac}$ field is about 1.2 meV  within the flake
which is much smaller than the energy gap within the green rectangle in Fig. \ref{wincy}. 
This allows us to treat the valence band as filled and frozen. For
the basis given by Eq. (\ref{basis}) we take up to 30 the lowest-energy states of the conduction band.

\begin{figure}\begin{tabular}{ll}
(a) &\includegraphics[width=0.8\columnwidth]{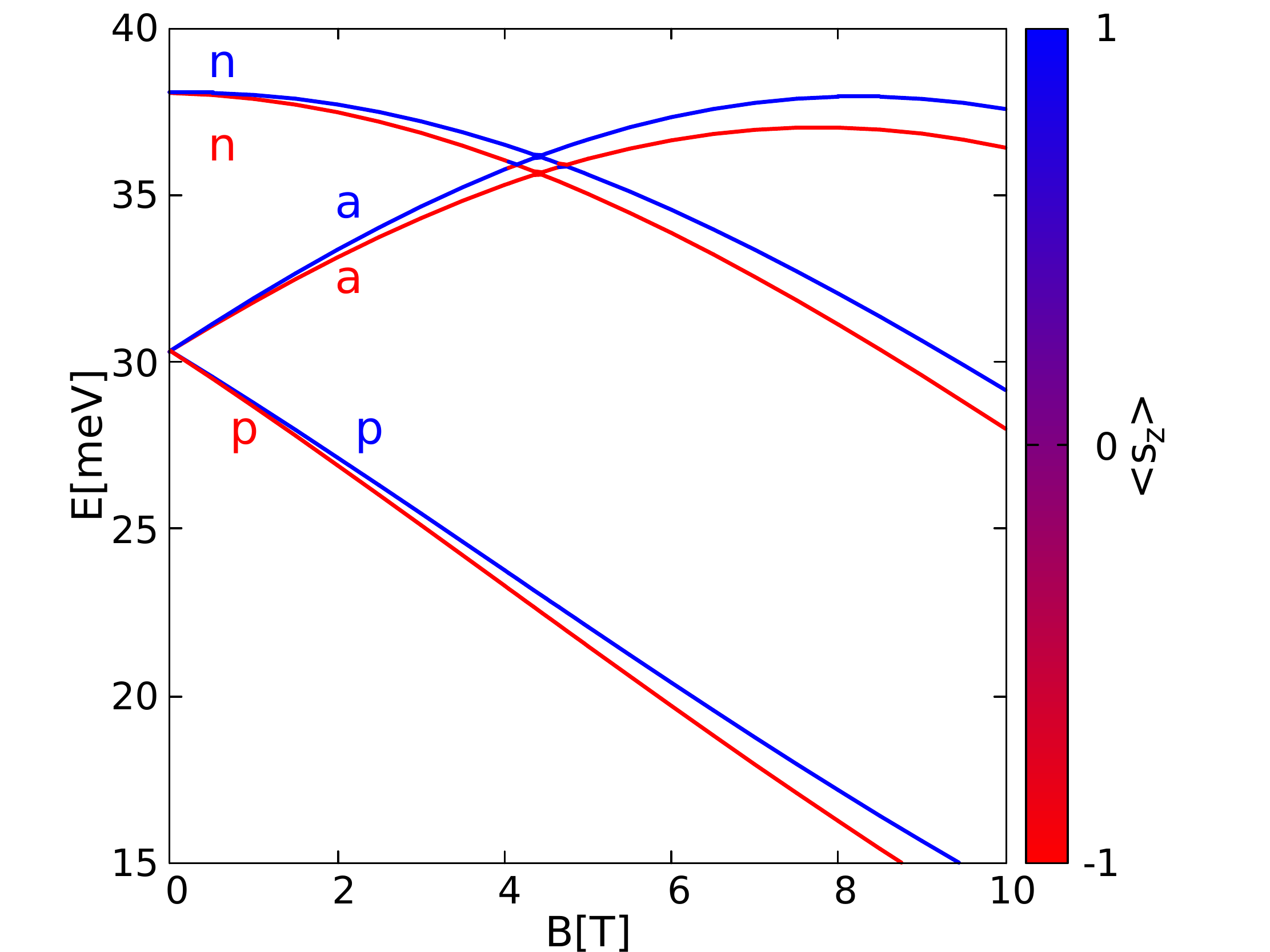} \\
(b) &\includegraphics[width=0.8\columnwidth]{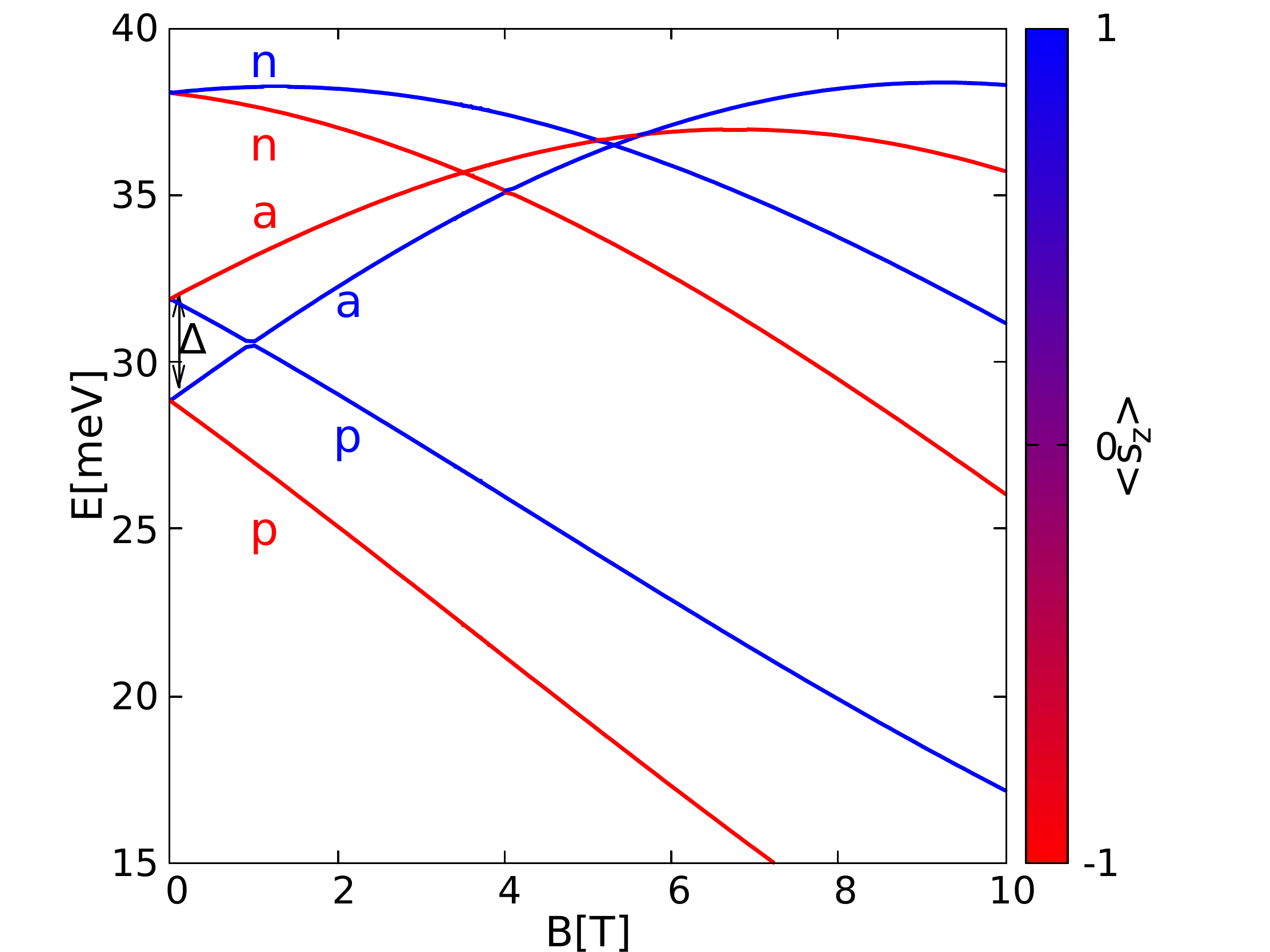} \\
(c) &\includegraphics[width=0.5\columnwidth]{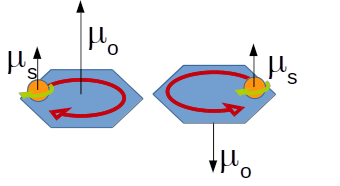} \\
\end{tabular}
  \caption{The lowest-energy conduction-band levels for the hexagonal silicene flake for $F_z=0$ without (a) and with the spin-orbit interactions (b). The colorscale
shows the average $z$ component of the spin in $\hbar/2$ units.
The plots (a,b) cover the region marked by the green lines in Fig. \ref{wincy} but with the Zeeman spin interaction included.
In (b) by $\Delta$ we mark the energy splitting induced by the spin-orbit interaction to the ground-state quadruplet of panel (a) at $B=0$. 
Levels marked by p,  a and $n$ indicate the magnetic moment generated by the electron currents which is parallel, antiparallel, and nearly absent, respectively.
(c) Schematics of the current circulation producing the orbital magnetic moment parallel (left) and antiparallel (right) to the spin magnetic moment.
For the spin-orbit coupling splitting at $B=0$ in plot (b) the ground-state (first excited) doublet corresponds to parallel (antiparallel) orbital $\mu_o$ and spin $\mu_s$ magnetic moments.
} \label{fz0}
\end{figure}

\section{Results and Discussion}

\begin{figure}\begin{tabular}{rlrl}
(a) &\includegraphics[width=0.4\columnwidth]{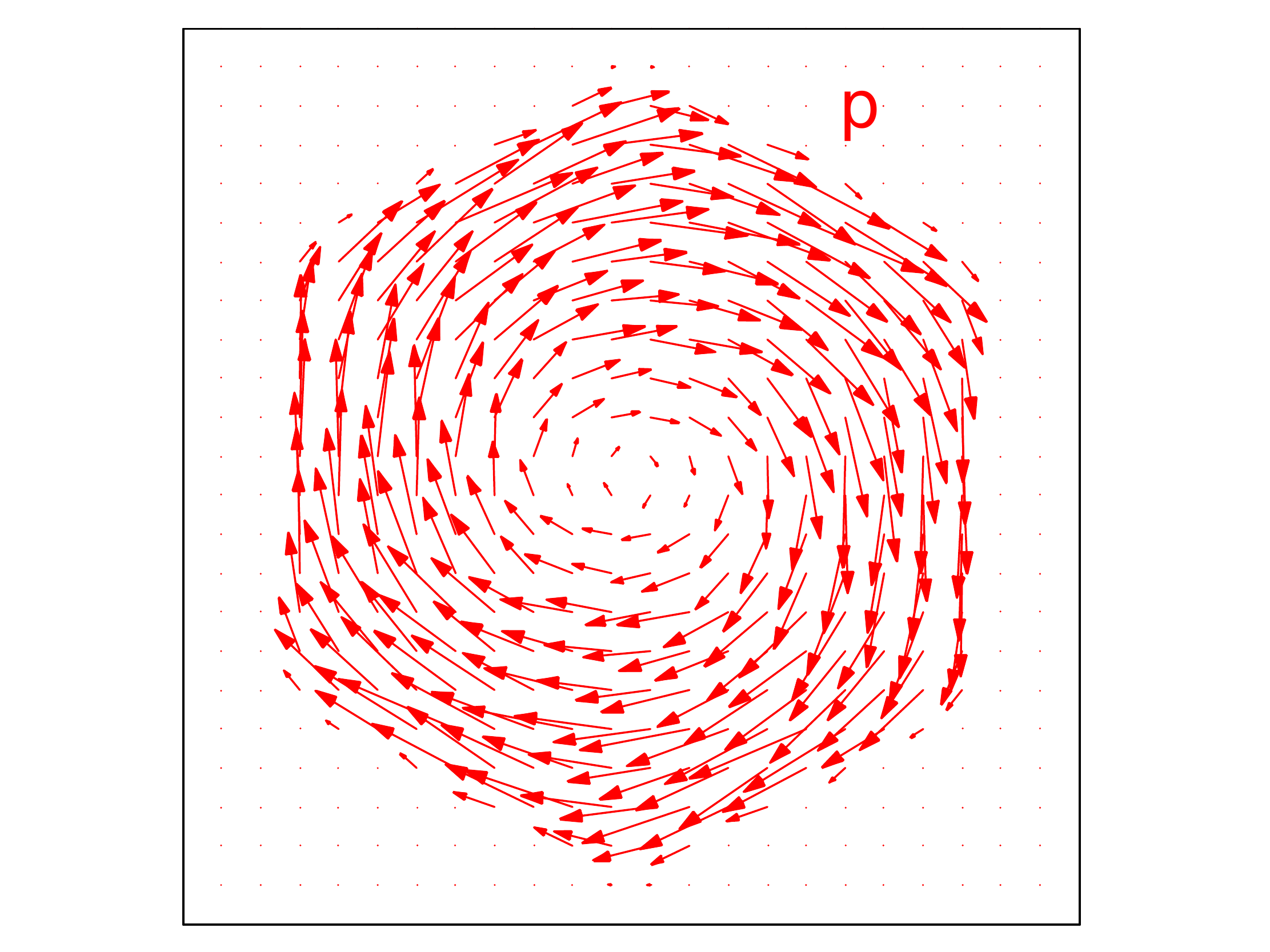} & (b) &\includegraphics[width=0.4\columnwidth]{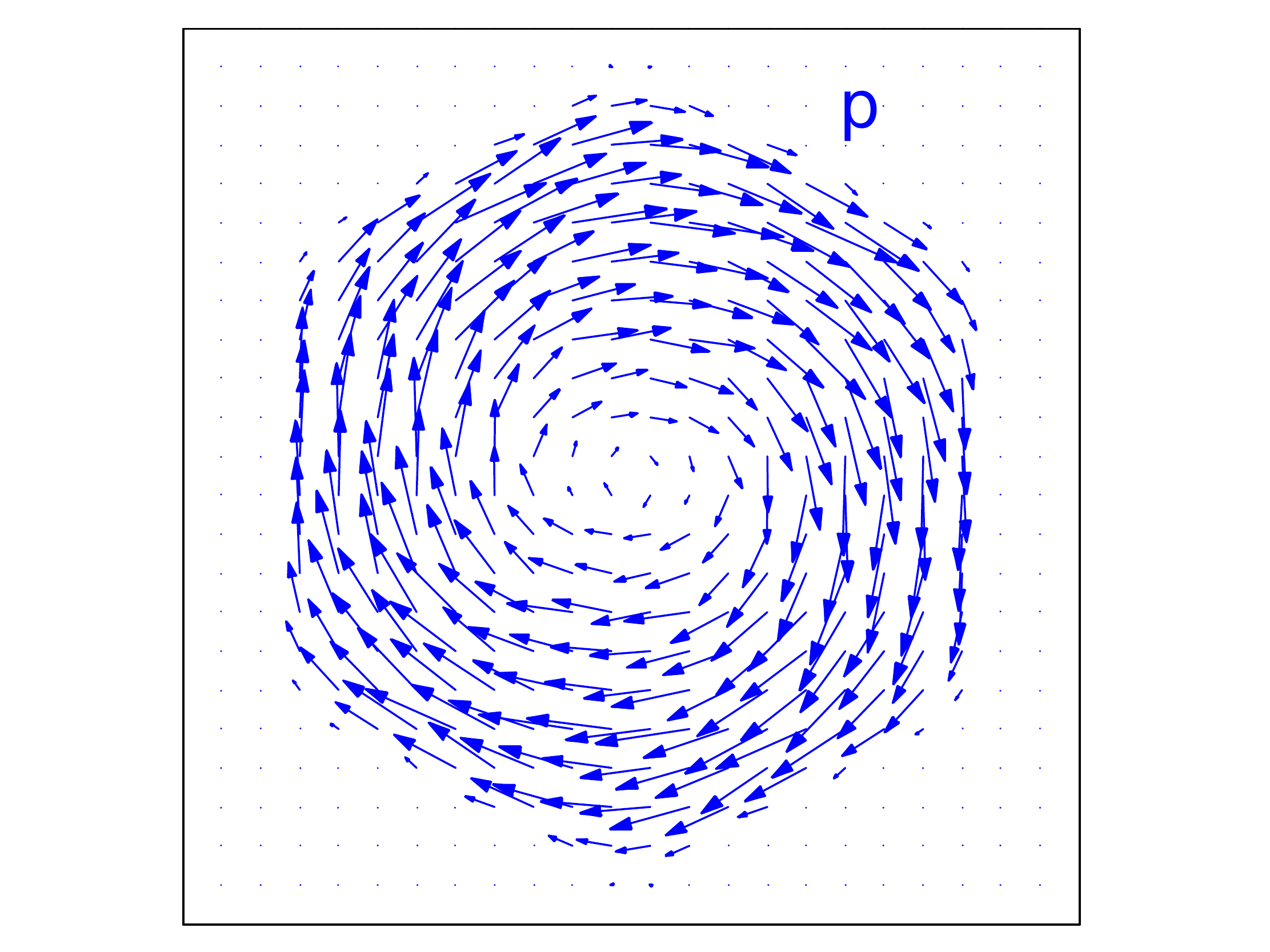} \\
(c) &\includegraphics[width=0.4\columnwidth]{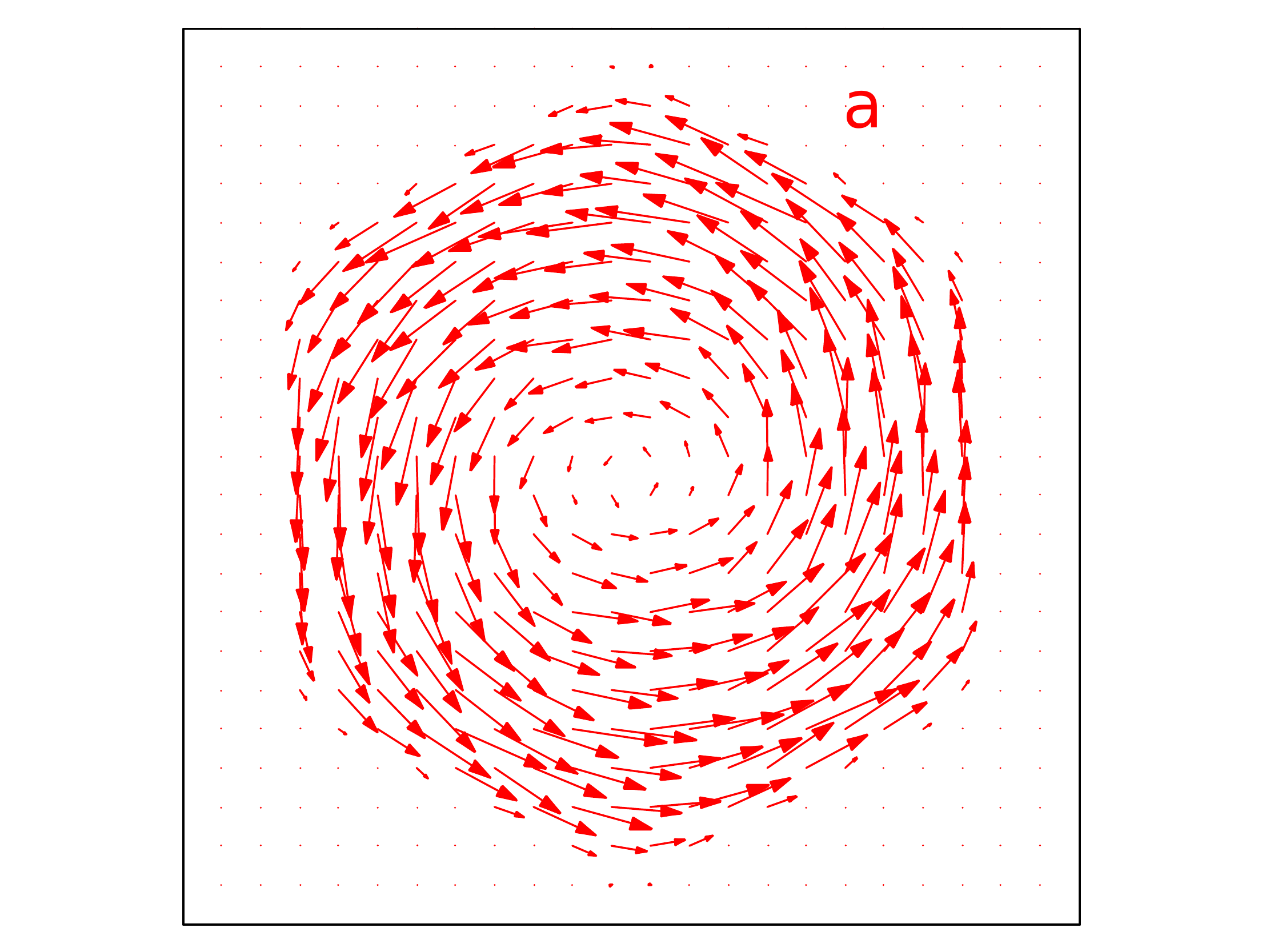} & (d) & \includegraphics[width=0.4\columnwidth]{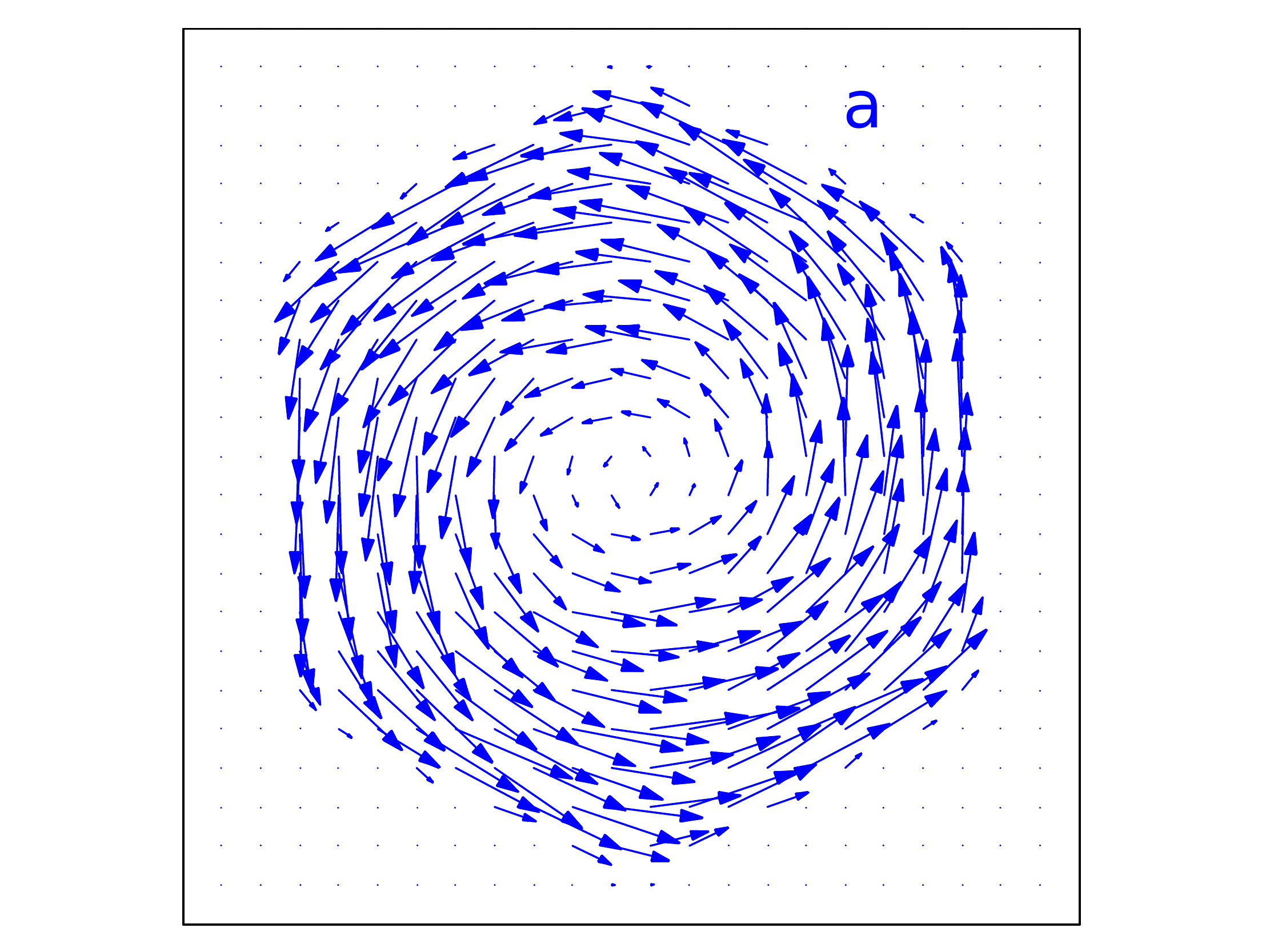}  \\
(e) &\includegraphics[width=0.4\columnwidth]{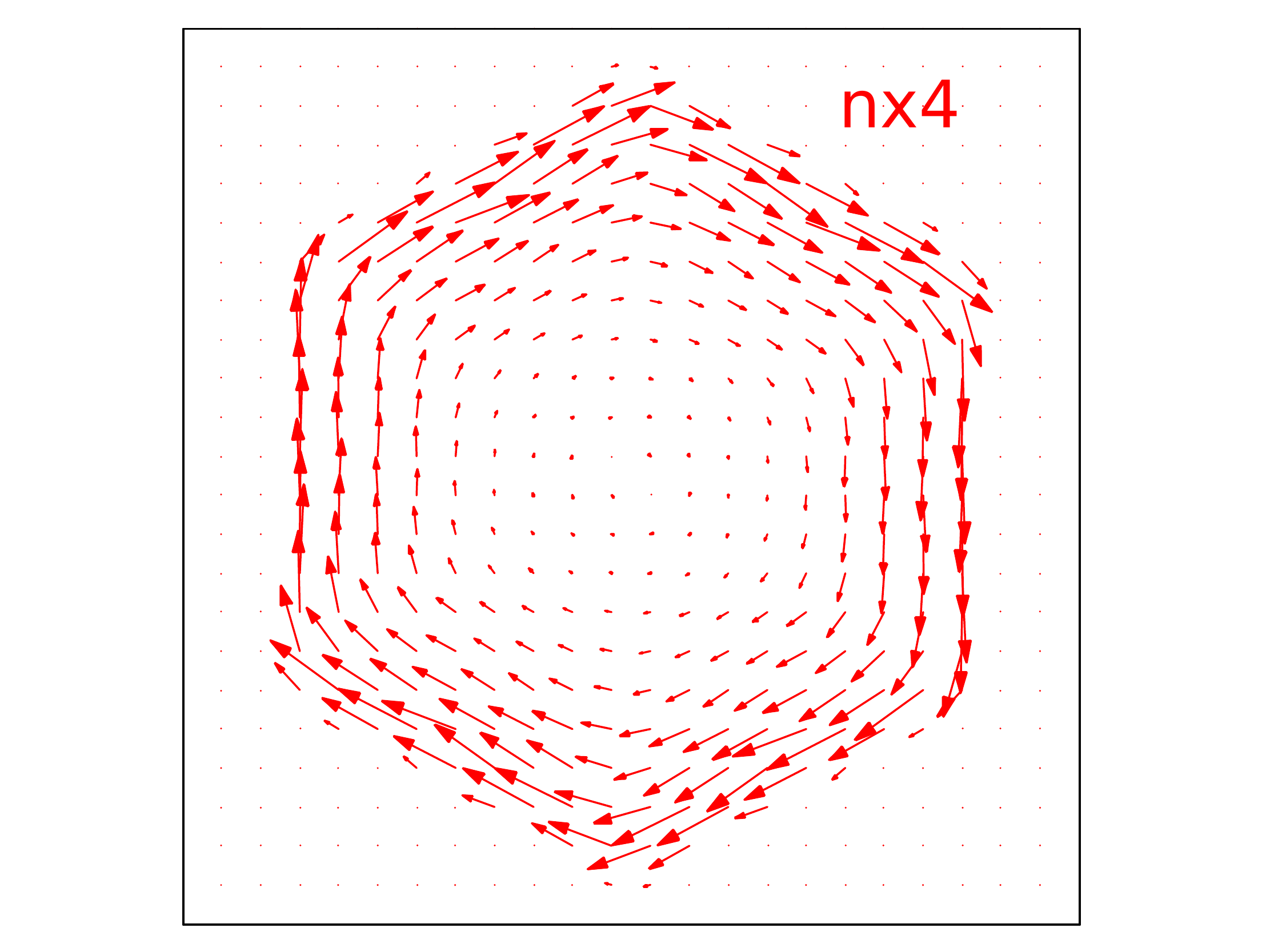} & (f) & \includegraphics[width=0.4\columnwidth]{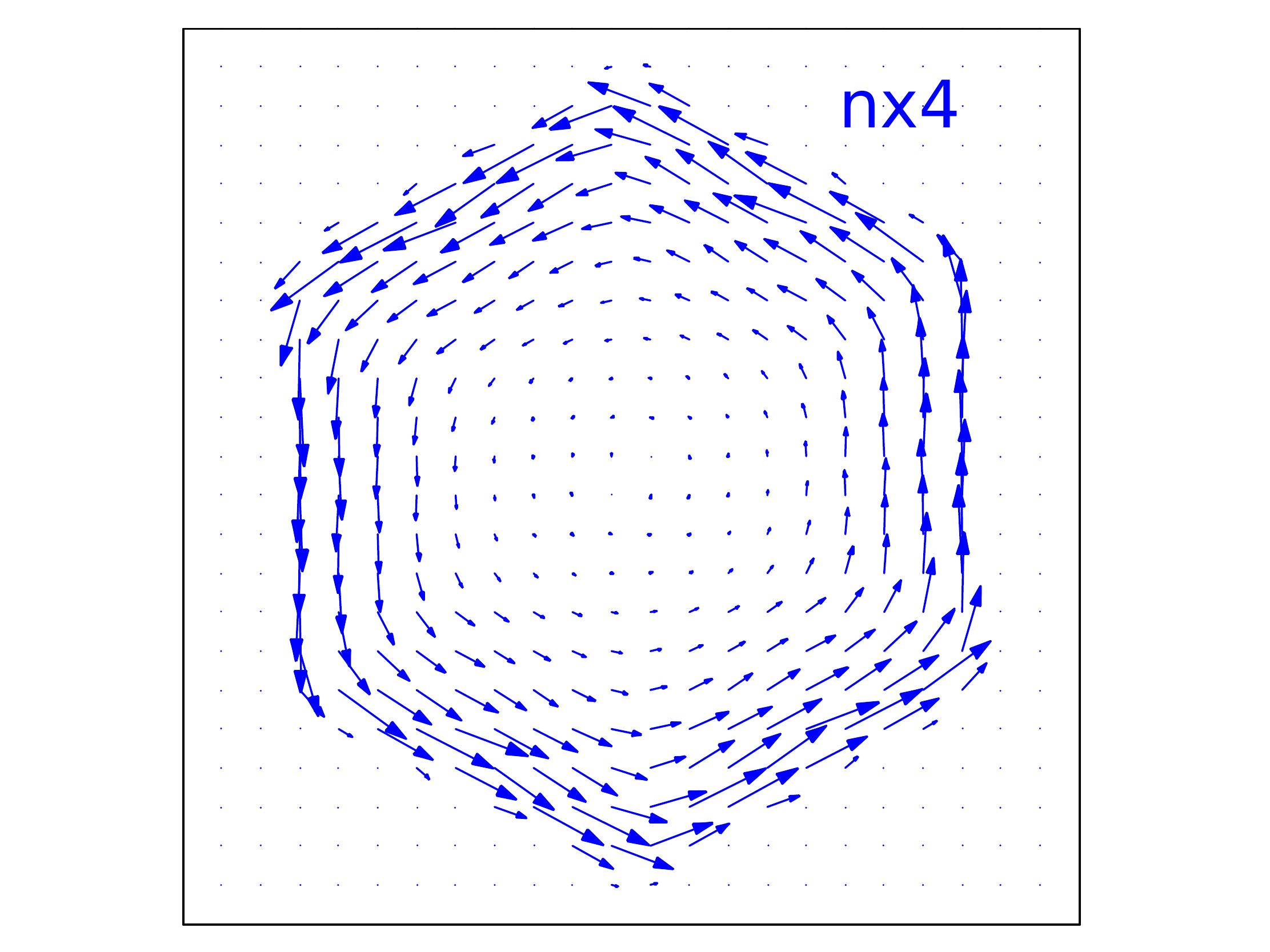} \\

\end{tabular}
  \caption{Electron current density  for $B=0.001$ T, $F_z=0$ with the spin-orbit coupling
[for the energy spectrum see Fig. \ref{fz0}(b)].
The left (a,c,e) [right(b,d,f)] column corresponds to the spin-down [spin-up] states. 
The energy levels are labeled by  p (a,b) and a (c,d) for the parallel or antiparallel
orientation of the magnetic dipole moment produced by the electron current with respect to the $z$ axis
as in Fig. \ref{fz0}. 
For the $p$ and a states the results in the absence of the spin-orbit interaction are nearly identical.
For the n states weak currents flow only in presence of the spin-orbit coupling. 
 Same scale of the current vectors is applied in the figure, with the exception of plots  for states labeled by $n$ (e,f), where the currents were multiplied by a factor of 4. 
}
 \label{pru}
\end{figure}

\subsection{Spectra in the absence of the vertical electric field}

The low-energy part of the conduction band spectrum for $F_z=0$ is given in Fig. \ref{fz0}.
Without the spin-orbit interaction [Fig. \ref{fz0}(a)] the structure of the lowest conduction-band energy levels  with the ground-state quadruplet and the excited doublet
is identical to the one obtained for the hexagonal armchair flakes of graphene --  see Fig. 5(a)  of Ref. \cite{zarenia,uwaga}.

When the spin-orbit coupling interactions are included [Fig. \ref{fz0}(b)] the ground-state quadruplet [Fig. \ref{fz0}(b)] is split to doublets 
separated by the energy of $\Delta\simeq 3.6$ meV [cf. Fig. \ref{fz0}(a) and Fig. \ref{fz0}(b)]. 
The Rashba spin-orbit terms are not resolved in the energy spectrum scale of Fig. \ref{fz0}(b) and the difference between Fig. \ref{fz0}(a) and Fig. \ref{fz0}(b) is entirely due to the intrinsic spin-orbit interaction. 
The first excited energy level is degenerate  only with respect to the spin.  

The armchair edge mixes the valleys in the Hamiltonian eigenstates \cite{zarenia} and the angular momentum for a hexagonal flake is not a good quantum number
(see Section \ref{dirac}).
Therefore,  we  refer to the separate states by the orientation of the magnetic dipole moments: the orbital moment 
generated by the electrical currents within the flake and by the spin. In labeling the states we take the values obtained at $B=0$ in the absence of the spin-orbit interactions.
Accordingly, the energy levels in Fig. \ref{fz0} and below are labeled by $p$, a and $n$, for the orbital magnetic moments ''parallel'', ''antiparallel''  to the external field or ''none'', respectively. 
 
Figure \ref{pru} shows the current distribution in the lowest six energy states.
The electron current \cite{waka} that flows from spin-orbital  $m\alpha$ to the spin-orbital  $n\beta$ in the tight-binding wave function 
$\Psi$ is calculated as ${{ j}}_{m\alpha n \beta}=\frac{i}{\hbar}\left(h_{m\alpha n\beta }\Psi_{m\alpha}^*\Psi_{n\beta}-h_{n\beta m\alpha}\Psi_{m\alpha}\Psi_{n\beta}^* \right)$. Figure \ref{pru} shows the currents in the majority spin component of the wave function. The other is generally negligible. Outside avoided crossings opened by the Rashba interaction (see below) the spin is nearly polarized in the $\pm z$ direction.

For the excited doublet a current circulation at $B=0$ is observed only when the spin-orbit interaction is present.  
The currents in the n states induced by the spin-orbit coupling  are weak compared to the ones that flow in a and $p$ states and require a scaling factor of 4 for presentation in Fig. \ref{pru}(e,f).

The magnetic dipole moment generated by the silicene flake is associated with the energy variation $\mu=-\frac{\partial E}{\partial B}$ \cite{szafran}, where $\mu=\mu_o+\mu_s$, $\mu_o$ is the orbital moment and  $\mu_s$ the spin  moment. The magnetic orbital moment results from the current circulation within the flake
$\mu_o=-\frac{e}{2} \int  ({\bf r}\times {\bf j}) { d^2r}$.
In the absence of the vertical electric field the orbital magnetic moment dominates over the spin moment. On the energy scale the latter   induces only a relatively small energy level splitting via the spin Zeeman interaction -- see the blue and red energy levels in Fig. \ref{fz0}(a).

Based on the above results we find that the splitting of the ground-state quadruplet to doublets separated by  $\Delta$ in Fig. \ref{fz0}(b)
shifts down (up) on the energy scale for states with the orbital magnetic dipole moment parallel (antiparallel) to the spin magnetic moment [see Fig. \ref{fz0}(c)].
 The energy of the n states with no 
  orbital magnetic moments  [Fig. \ref{fz0}(a)]  
is only weakly changed when  the spin-orbit interactions are included [cf. Fig. \ref{fz0}(a) and (b)]. 
 The spin-orbit interaction introduces a built-in magnetic field which for $n$ energy levels shifts the $E(B)$ extrema off $B=0$ [Fig. \ref{fz0}(b)]
and the current circulation in these states -- although weak -- no longer vanishes at $B=0$ -- see Fig. \ref{pru}(e,f).


\begin{table*}[htbp]
$F_z=0$, $B=1$ T,  p$\downarrow$ as the initial state \\
\begin{tabular}{|c|c|c|c|c|c|}\hline
final state & $|\langle x \rangle_{A\uparrow}|$ & $|\langle x\rangle_{A\downarrow}|$ & $|\langle x\rangle_{
B\uparrow}|$ & $|\langle x\rangle_{B\downarrow}|$ & $|\langle x\rangle|$  \\ \hline
p$\downarrow\longrightarrow$ n$\uparrow $ &  $5.290\times 10^{-4}(+)$  & $5.611\times 10^{-4} (-)$ &    $5.290\times 10^{-4}(+)$  & $5.611\times 10^{-4} (-)$ & $6.435\times 10^{-5}$ \\ 
p$\downarrow\longrightarrow$ n$\downarrow $ &  $4.543\times 10^{-7}(+)$  & $3.214 (+)$ &    $4.543\times 10^{-7}(+)$ &  $3.124(+)$ & 6.429  \\

p$\downarrow\longrightarrow$ a$\downarrow$ & $2.864\times 10^{-7}(+)$  &$2.438(-)$&  $2.864\times 10^{-7}(-)$ & $2.438(+)$ &   0 \\

p$\downarrow\longrightarrow$ p$\uparrow$ & $2.179\times 10^{-4}(+)$ & $2.398\times 10^{-4}(+)$ & $2.179\times 10^{-4}(-)$ &  $2.398\times 10^{-4}(-)$ &0 \\

p$\downarrow\longrightarrow$ a$\uparrow$ & $3.812\times 10^{-9}(+)$  &$3.133\times 10^{-9}(-)$&  $3.812\times 10^{-9}(-)$ & $3.133\times 10^{-9}(+)$ &   0 \\

\hline
\end{tabular} 
\vspace{0.5cm}\\
$F_z=0.25$ V/\AA, $B=1$ T,  p$\downarrow$ as the initial state \\
\begin{tabular}{|c|c|c|c|c|c|}\hline
final state & $|\langle x \rangle_{A\uparrow}|$ & $|\langle x\rangle_{A\downarrow}|$ & $|\langle x\rangle_{
B\uparrow}|$ & $|\langle x\rangle_{B\downarrow}|$ & $|\langle x\rangle|$  \\ \hline
p$\downarrow\longrightarrow$ n$\uparrow $ &  $2.610\times 10^{-2}(+)$  & $3.857\times 10^{-3} (-)$ &    $2.237\times 10^{-3}(-)$  & $2.6277\times 10^{-3} (+)$ & $1.208\times 10^{-2}$ \\ 

p$\downarrow\longrightarrow$ n$\downarrow $ &  $3.865\times 10^{-4}(+)$  & $5.888 (+)$ &    $2.684\times 10^{-5}(+)$ &  $0.439(+)$ & 6.328  \\

p$\downarrow\longrightarrow$ a$\downarrow$ & $2.529\times 10^{-4}(+)$  &$4.550(+)$&  $1.781\times 10^{-5}(-)$ & $0.286(-)$ &   $4.265$ \\

p$\downarrow\longrightarrow$ p$\uparrow$ & $3.487\times 10^{-2}(+)$ & $6.034\times 10^{-2}(-)$ & $1.997\times 10^{-3}(+)$ &  $3.949\times 10^{-4}(-)$ & $2.741\times 10^{-2}$  \\

p$\downarrow\longrightarrow$ a$\uparrow$ & $1.485\times 10^{-7}(+)$  &$1.621\times 10^{-7}(-)$&  $8.253\times 10^{-9}(+)$ & $9.801\times 10^{-9}(-)$ &   $1.516 \times 10^{-7}$ \\

\hline
\end{tabular} 
\caption{The absolute value of the dipole matrix elements $|\langle x\rangle|$ in nanometers (last column) between the p$\downarrow$ ground state and the excited states listed in the first column.  The columns from the second to the fifth give the absolute values of the contributions to the matrix element for a given (A/B) sublattice and the spin-component ($\uparrow\downarrow$). The $\pm$ sign in the parentheses denotes the sign of the contribution taken from its real part. The upper and lower tables correspond to $F_z=0$ and $F_z=0.25$ V/\AA, respectively.
}
\end{table*}

\begin{table}

\begin{tabular}{|c|c|c|c|}\hline
$F_z$ [V/\AA] & (a,p)($\uparrow\downarrow$) & n ($\uparrow\downarrow$) & $E_g$ [meV] \\ 
\hline
0 & 0.5 & 0.5 & 57.7 \\
0.125 & 0.843 & 0.800 & 80.5 \\
0.25 & 0.941 & 0.916 &  126.8 \\
0.5 & 0.982 & 0.974 & 234.4 \\
1 & 0.995 & 0.993 & 460.2 \\
1.5 & 0.997 & 0.996 & 688.5 \\ \hline
\end{tabular} 
\caption{Second and third column: the part of the electron density on the $A$ sublattice for varied $F_z$  (first column)
for the states of the $a,p$ $(\uparrow\downarrow)$ quadruple and $n(\uparrow\downarrow)$ doublet. The last column contains the energy gap $E_g$, i.e. the spacing between the lowest conduction band and highest valence band levels. 
The perpendicular magnetic field of $B=0.001$ T was applied. 
}
\end{table}

\subsection{Transition matrix elements}

The rate of the resonant spin flips that can be achieved by the AC electric fields is determined 
by the transition matrix elements $\langle \Psi_i |F_{ac} x |\Psi_f \rangle$,
where $\Psi_i$ and $\Psi_f$ are the wave functions of the initial and final states. 
Table I lists the absolute values of the matrix elements $|\langle \Psi_i | x |\Psi_f\rangle|$ for the ground state set as the initial one $i=p\downarrow$.
The initial- and the final-state wave functions can be expressed by contributions of separate sublattices and the spin components
$\Psi=\Psi^{A\uparrow}+\Psi^{B\uparrow}+\Psi^{A\downarrow}+\Psi^{B\downarrow}$.
The $x$ operator is spin- and sublattice-diagonal, hence
$\langle \Psi_i | x |\Psi_f\rangle=\langle \Psi_i^{A\uparrow}|x |\Psi_f^{A\uparrow}\rangle 
+\langle \Psi_i^{A\downarrow}|x| \Psi_f^{A\downarrow}\rangle
+\langle \Psi_i^{B\uparrow}|x |\Psi_f^{B\uparrow}\rangle 
+\langle \Psi_i^{B\downarrow}|x| \Psi_f^{B\downarrow}\rangle\equiv \langle x _{A\uparrow}\rangle
+\langle x _{A\downarrow}\rangle
+\langle x _{B\uparrow}\rangle 
+\langle x _{B\downarrow}\rangle$.
 The columns in Table I from the 2nd to the 5th contain the absolute values of the contributions to the matrix elements from the A and B sublattices and the spin components of the integrands.
Let us first focus on the upper part of Table I that corresponds to $F_z=0$. 
We can see that all the transitions within the ground-state quadruple -- with spin flip or not -- are forbidden. Only the transitions from p$\downarrow$ to n$\uparrow\downarrow$ states can be induced by the AC electric field,
and the rate of the transition with the spin inversion p$\downarrow\longrightarrow$n$\uparrow$ is by five orders of magnitude slower than the spin-conserving transition p$\downarrow\longrightarrow$n$\downarrow$. 

Let us look at the contributions over the spins and sublattices for $F_z=0$. 
For the final state $f=$a$\uparrow$ the values of all four separate contributions are negligibly small, of
the order of $10^{-9}$ nm. For $f=$p$\uparrow$ the contributions are larger --  of the order of $10^{-4}$ nm.
Note that the non-zero values of the separate contributions between the states of opposite spins are entirely due to the Rashba interactions which are non-diagonal 
in $s_z$. 
 However, for $f=$p$\uparrow$ the contributions
from one sublattice are exactly cancelled by the contribution of the other sublattice (upper part of Table I).
For the spin-conserving transition to a$\downarrow$ the separate contributions are large but
again cancel over the sublattices. 

The vertical electric field $F_z$ lifts the cancellation of the contributions from separate sublattices. The vertical field distinguishes the  sublattices and
for $F_z>0$ the field shifts most of the wave function to the $A$ sublattice (see Table II). The lower part of Table I indicates the changes to the 
matrix elements introduced by nonzero $F_z$. The spin-flipping transition within the ground-state quadruple p$\downarrow\longrightarrow$p$\uparrow$ has now only three orders of magnitude lower matrix element  than the spin-conserving one. The p$\downarrow\longrightarrow$a$\uparrow$ spin-flipping transition has still a negligibly small rate since already the contributions of the sublattices are small.

\subsection{Energy spectra in the vertical electric field}
\begin{figure}
\begin{tabular}{l}
(a) \includegraphics[width=0.45\columnwidth]{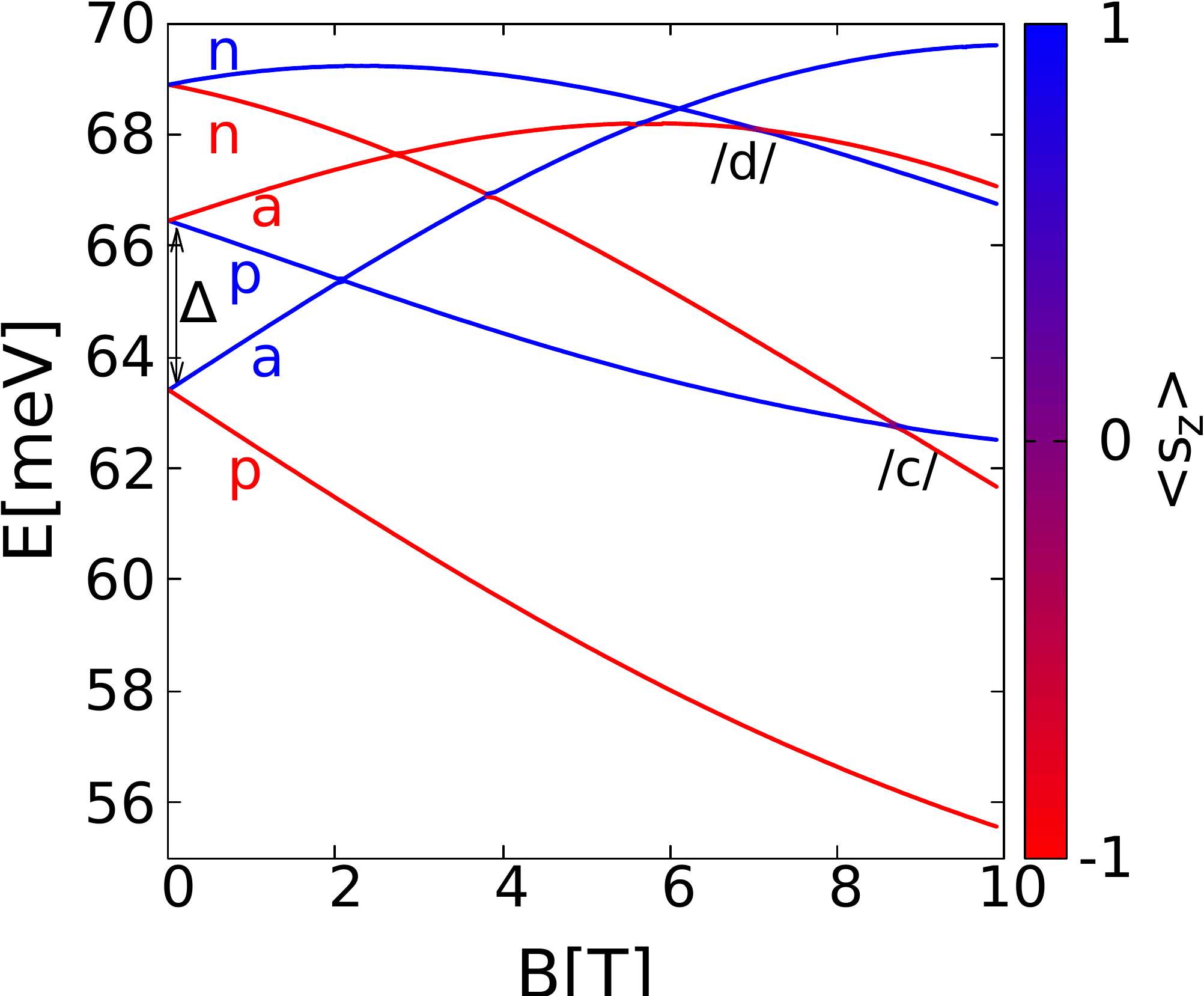}  (c) \includegraphics[width=0.42\columnwidth]{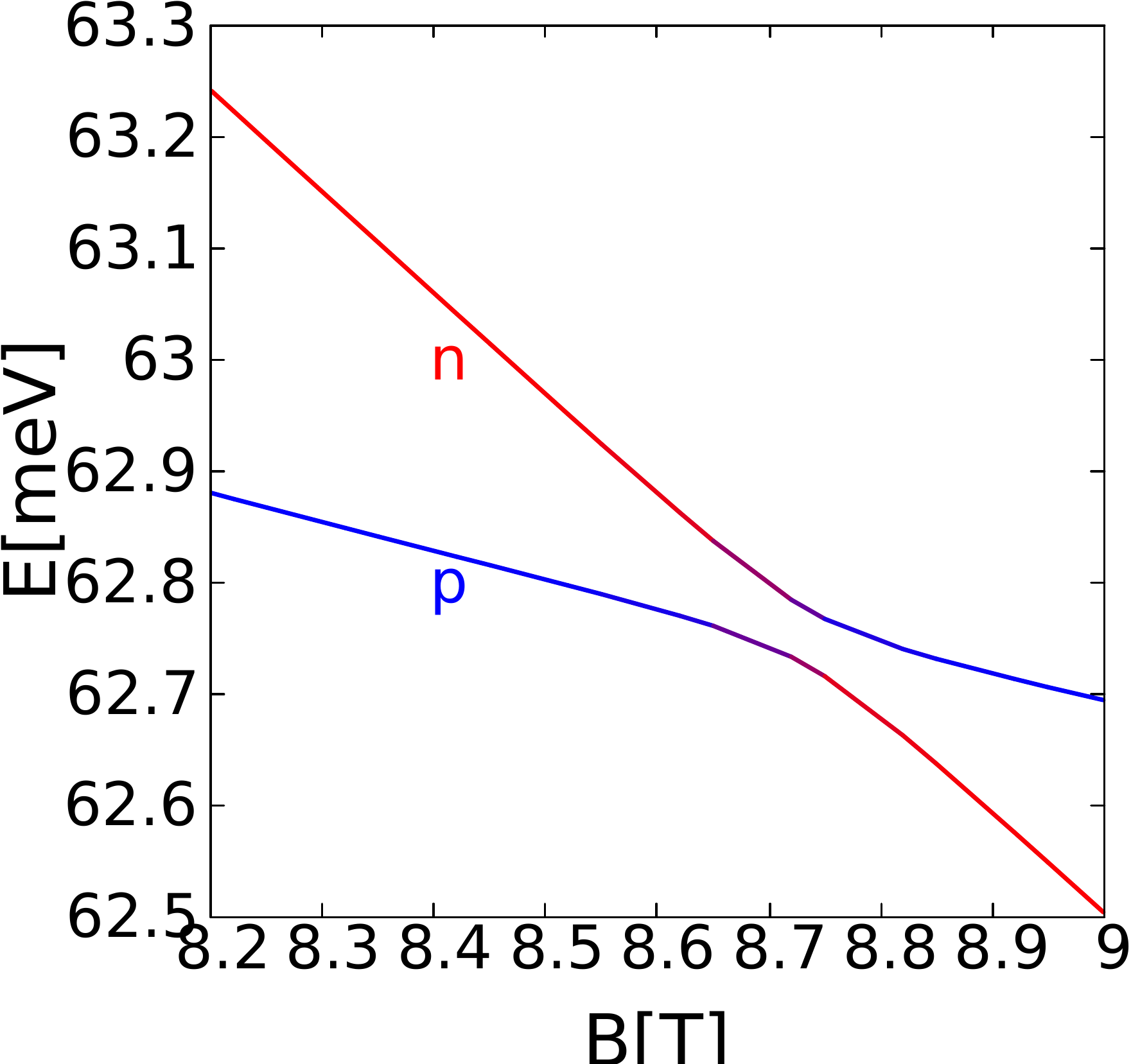} \\

(b) \includegraphics[width=0.45\columnwidth]{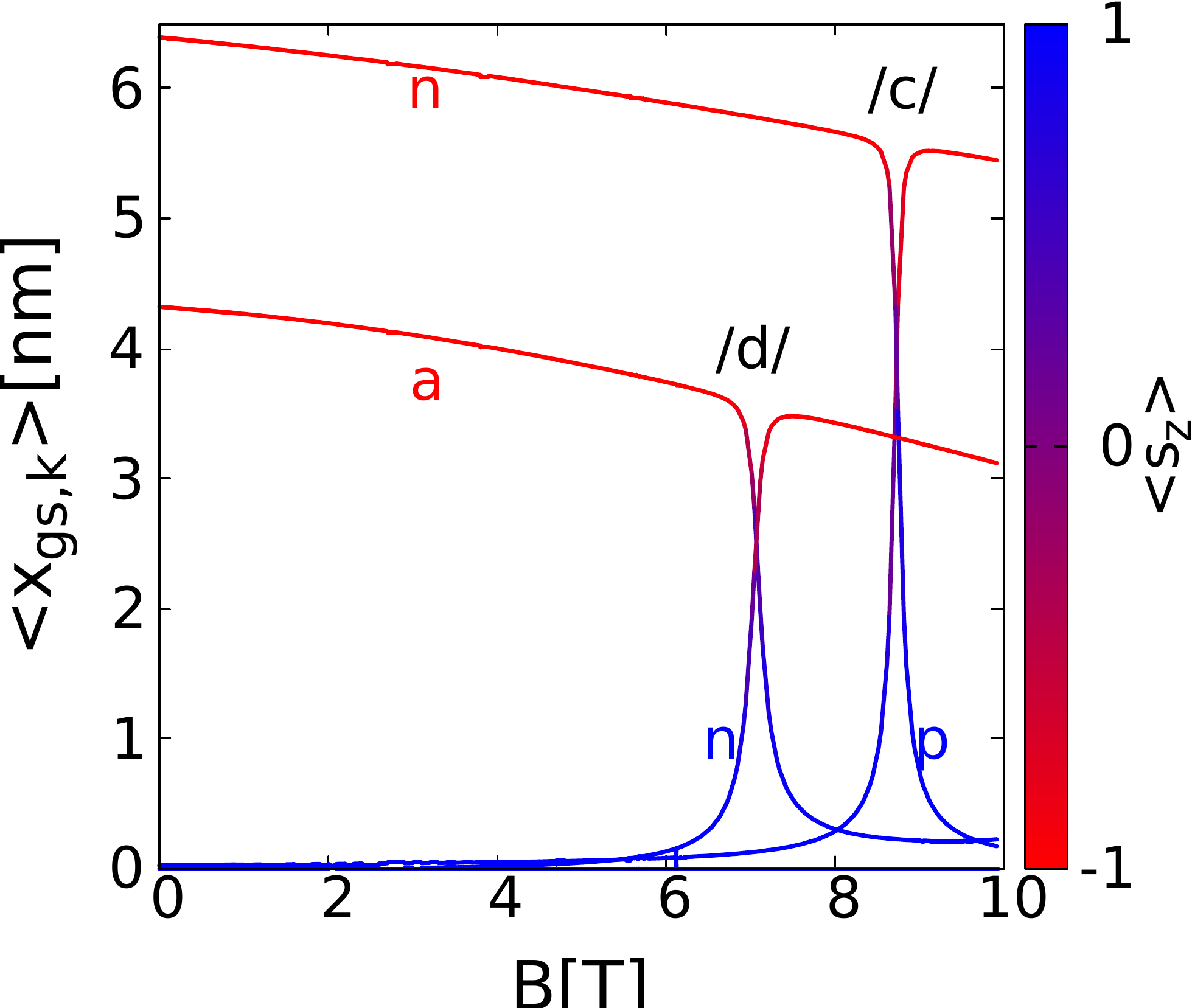}  
 (d) \includegraphics[width=0.42\columnwidth]{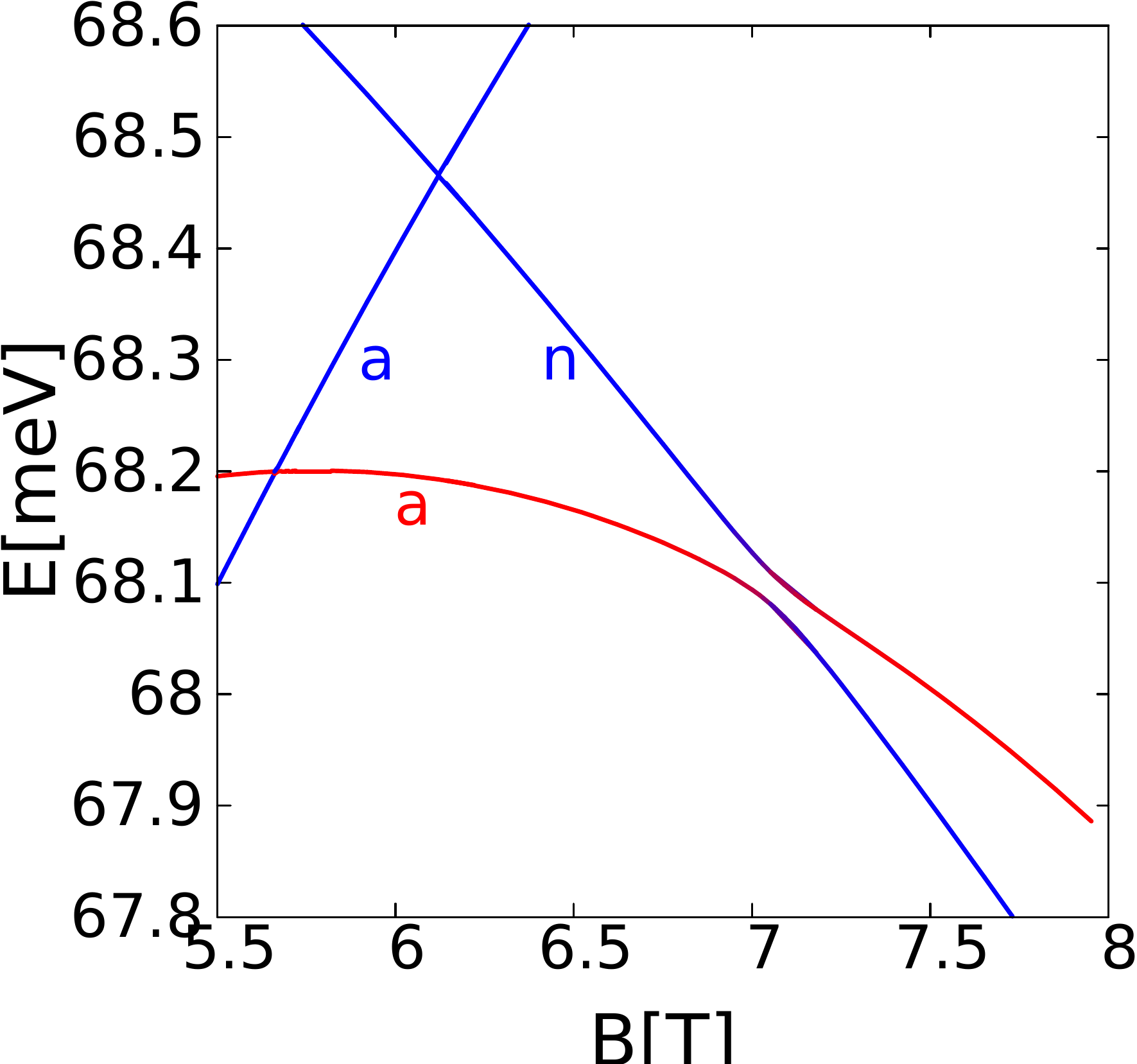} \\
\end{tabular}
  \caption{(a) Energy spectrum for the silicene flake in the vertical electric field of $F_z=0.25$ V/\AA . The spin-orbit interaction and the Zeeman effect are included. 
(b) The transition dipole matrix elements between the p$\downarrow$ ground-state and the excited states.
(c) and (d) show the enlarged fragments of (a) for the avoided crossings that are marked by /c/ and /d/ respectively in plots (a) and (b).
The spin-flipping transitions acquire large matrix elements near the avoided crossings 
opened by the Rashba interaction between the energy levels of opposite spin: n$\downarrow$ and p$\uparrow$ (c)  and n$\uparrow$ and a$\downarrow$ (d).
 The  colorscale shows the average $z$ component
of the spin in $\hbar/2$ units. In (b) the spin of the final state is marked with the color.
} \label{fk25}
\end{figure}

\begin{figure}
\begin{tabular}{ll}
(a) &\includegraphics[width=0.6\columnwidth]{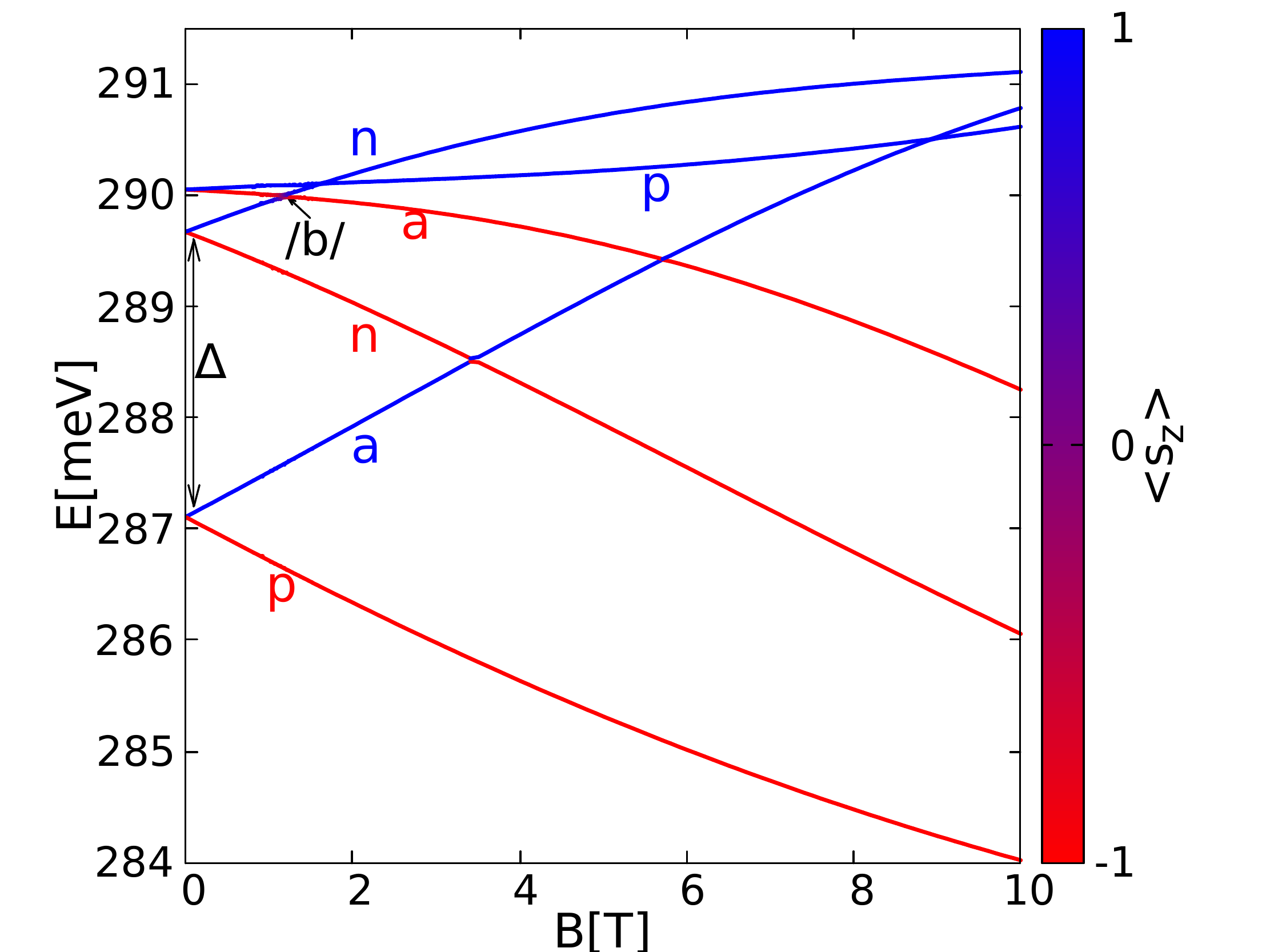} \\
(b) &\includegraphics[width=0.6\columnwidth]{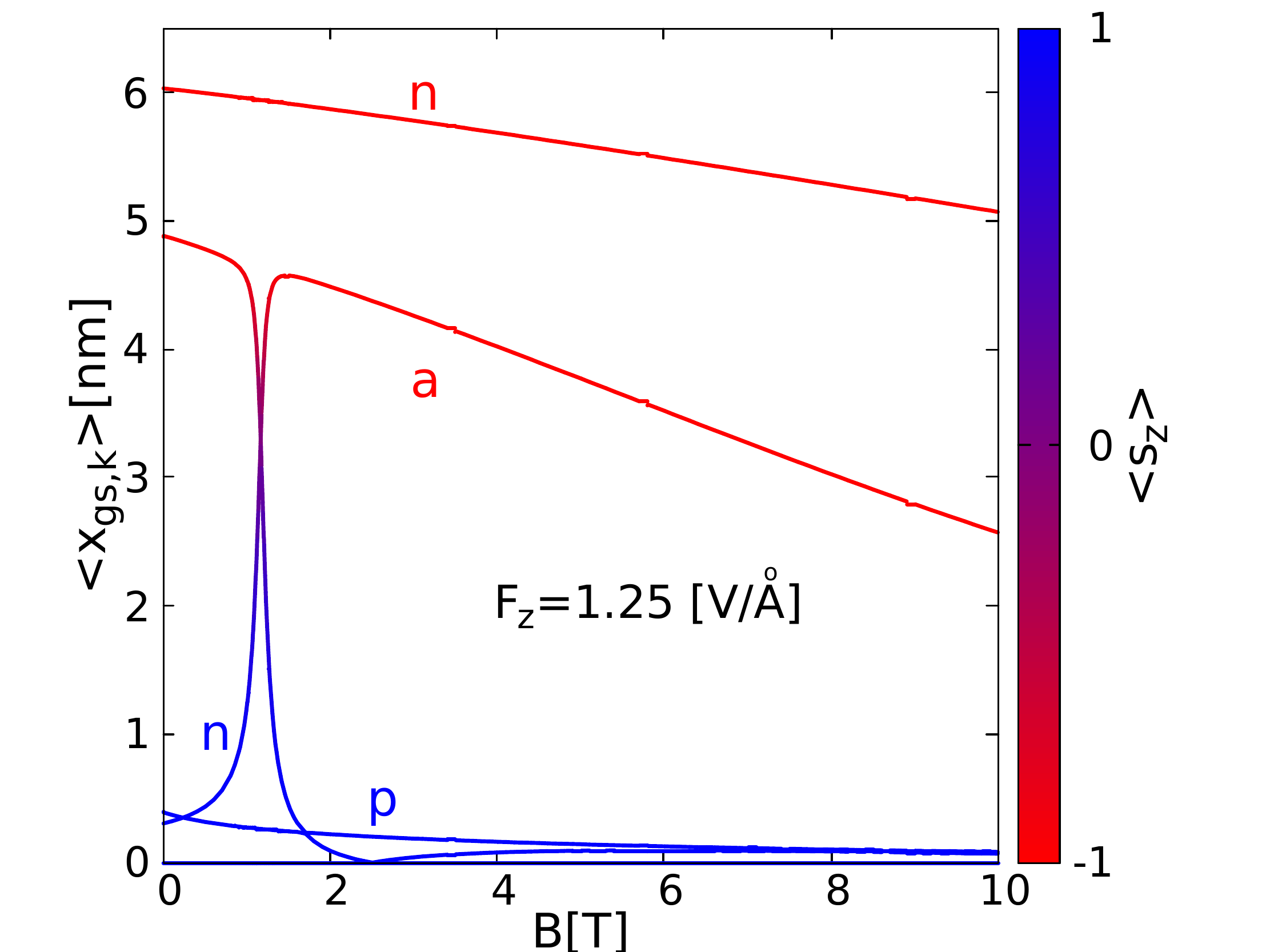}\\
 \end{tabular}
  \caption{(a) Energy spectrum for the silicene flake for $F_z=1.25$V/\AA\  in presence of spin-orbit and Zeeman interactions.
 $\Delta$ is  the energy splitting induced by the spin-orbit interaction. 
(b) The transition dipole matrix elements between the p$\downarrow$ ground-state and the excited states. The position of this avoided crossing is marked with /b/ in (a).} \label{f1k25}
\end{figure}

The current flow, at the atomic scale, occurs along the bonds between the sublattices \cite{gru}.
Therefore,  localization of the wave function on the single sublattice [Table II]
hampers the current circulation within the flake.  
The vertical electric field quenches the currents, the orbital magnetic dipole moment is reduced,
and so is the  scale of the variation of energy levels with $B$ -- compare Fig. \ref{fz0}(a) for $F_z=0$ with Fig. \ref{fk25}(a) for $F_z=0.25$ V/\AA\; and Fig. \ref{f1k25}(a) for $F_z=1.25$ V/\AA.
In presence of the vertical electric field, 
 avoided crossings of states with opposite spin orientations can be resolved.
Figure \ref{fk25} contains signature of energy levels repulsion between
p$\uparrow$ and n$\downarrow$  levels [Fig. \ref{fk25}(c)]
as well as between a$\downarrow$ and n$\uparrow$  [Fig. \ref{fk25}(d)] levels.
The avoided crossings involve states of opposite spins but the same current orientation at $B=0$ [cf. Fig \ref{pru}]. Near the avoided crossings [Fig. \ref{fk25}(b)] the transition matrix element
from the ground-state to the states of opposite spin is radically increased.
The dependence of the matrix elements on  $F_z$ is therefore  nonlinear and nonmonotonic.

For $F_z=1.25$ V/\AA\; [see Fig. \ref{f1k25}(a)] only a single avoided crossing is observed in the spectrum -- when the spin-up $n$ energy level changes in order with spin-down a energy level.
The other crossing no longer occurs, since the p energy level  at $B=0$ lies higher than the n levels. 
In contrast to the results with $F_z=0$ [Fig. \ref{fz0}] in Fig. \ref{f1k25}(a) the main source of the energy variation is the spin Zeeman interaction, and the orbital magnetic moments are very small due to the current quenching noticed above. In this sense, the vertical electric field controls the magnetic properties of the flake as the source of the magnetic dipole moment. 

\begin{figure}
\begin{tabular}{ll}
(a) & \includegraphics[width=.7\columnwidth]{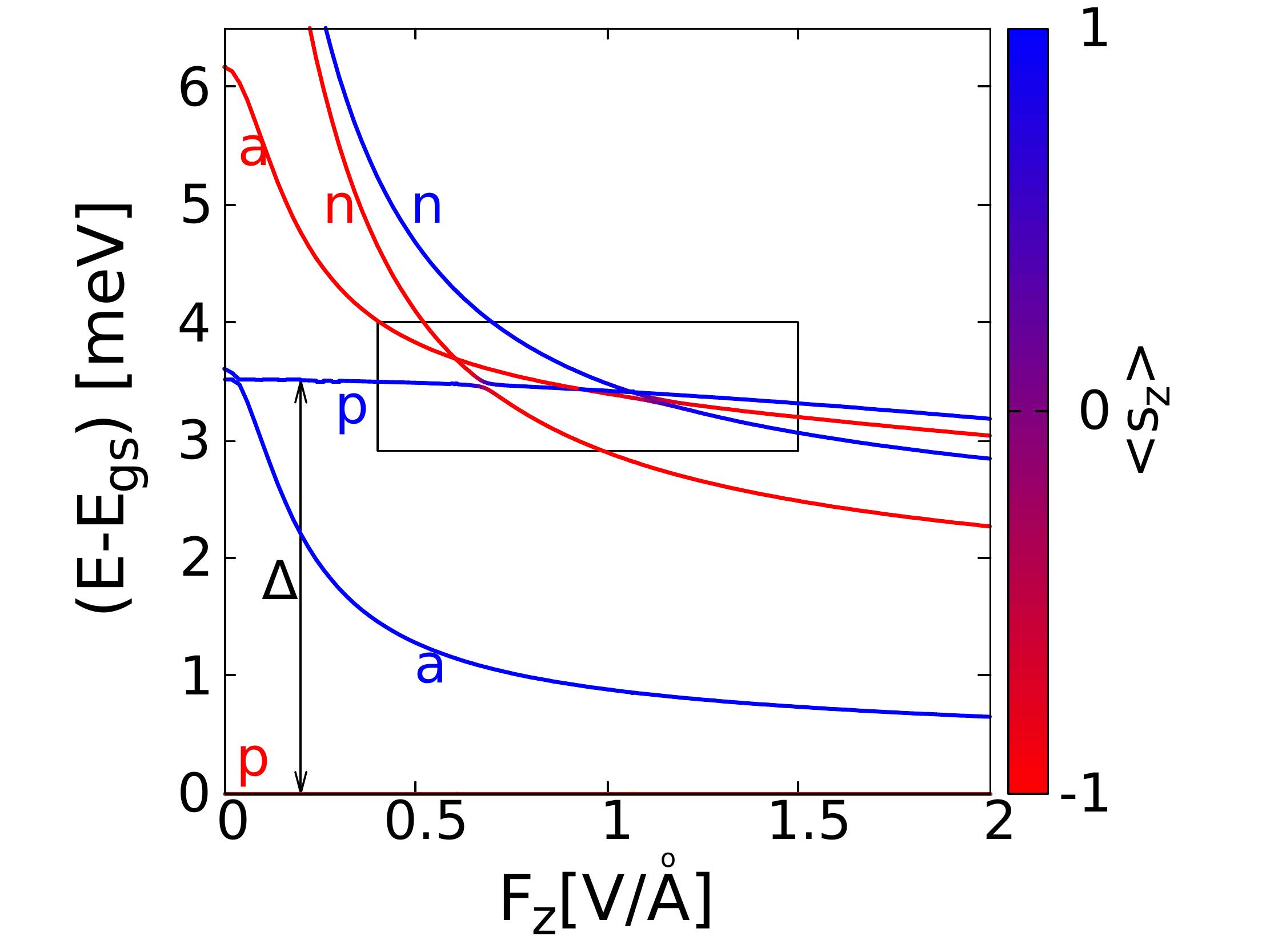} \\
(b) & \includegraphics[width=.7\columnwidth]{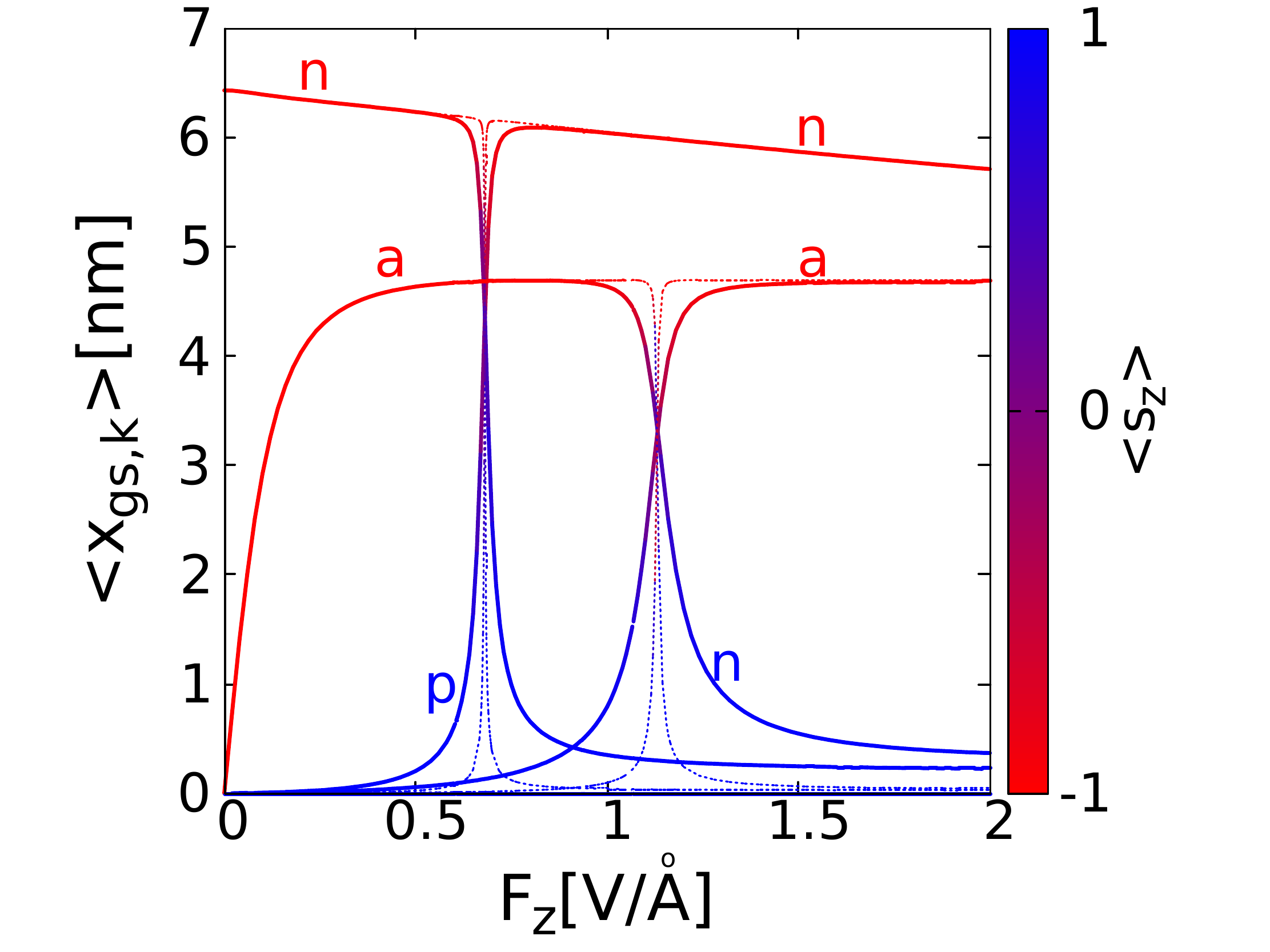} \\ 
(c) & \includegraphics[width=.7\columnwidth]{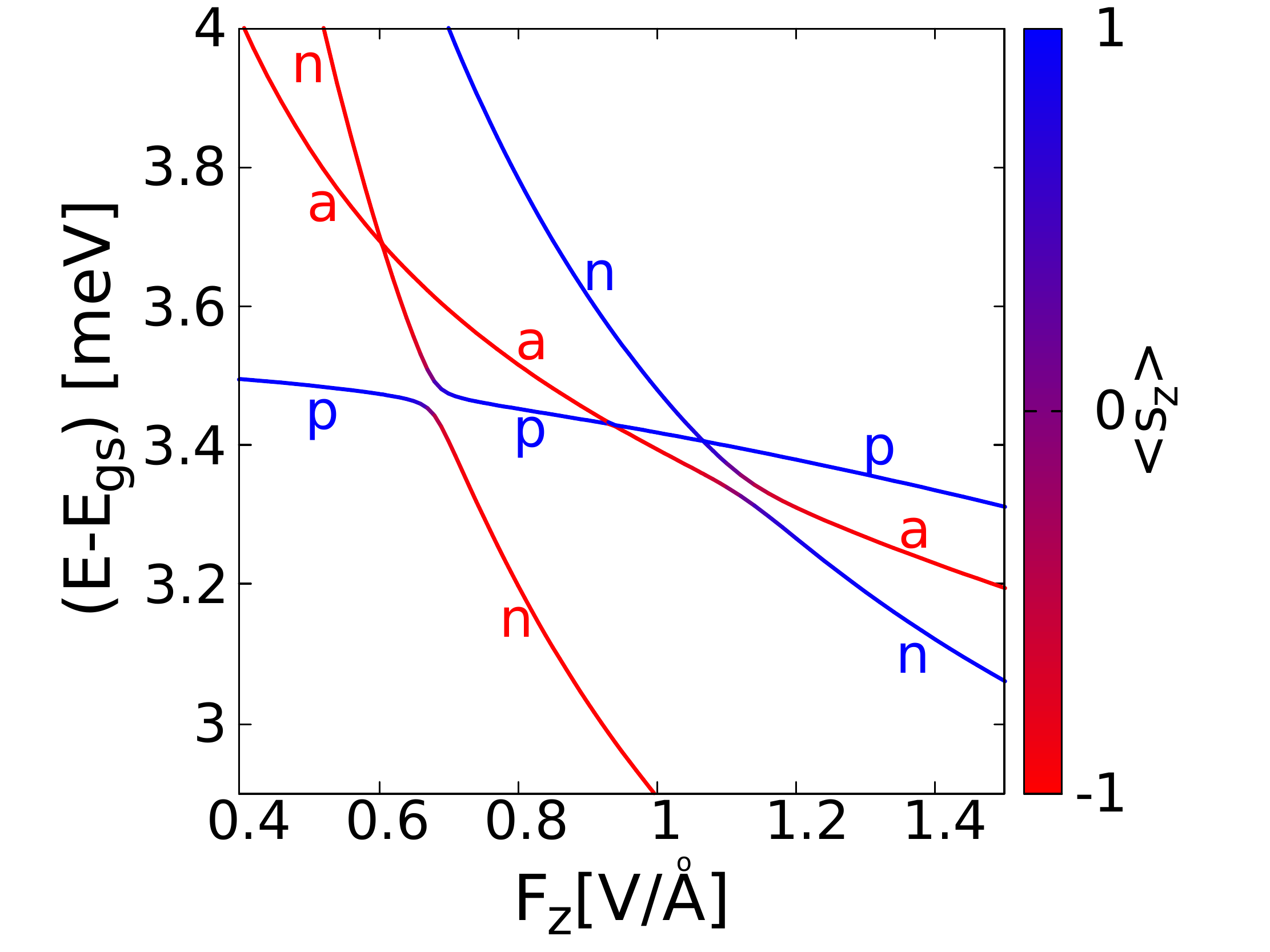} \\ 
\end{tabular}
  \caption{(a) The energy levels for $B=1$ T as functions of the vertical electric field in presence of spin-orbit and Zeeman interactions. $\Delta$ is the spin-orbit coupling splitting energy
for the ground-state quadruplet -- see Fig. \ref{fk25} and Fig. \ref{fz0}(b). The area marked by the rectangle is enlarged in (c).
(b) The transition matrix element between the p$\downarrow$ ground state and the excited states.
The thin  lines indicate the results for the Rashba parameter due to the external field set to zero $t_3=0$. }
 \label{wpulu}
\end{figure}

The dependence of the spectrum on the vertical field is summarized for $B=1$ T  in Fig.  \ref{wpulu}. 
For the vertical electric potential the energy reference level is set in the center between the $z$ positions of $A$ and $B$ sublattices. 
The energy gap as a function of $F_z$ is given in the last column of Table II. The energy of the lowest-conduction band states is  equal to half the energy gap. 
The gap grows with $F_z$ since 
the conduction (valence) band states get localized at the sublattice whose electrostatic potential
is increased (decreased) by the field. The growth of the n states energy is weaker 
since their localization on the A sublattice with growing $F_z$ is delayed [Table II].
 As a result, the n states enter in between the a and p states for $F_z>1$ V/\AA. Note that the spin-orbit energy level splitting $\Delta$ between p$\downarrow$ and p$\uparrow$ states only
weakly depends on the external electric field.

\begin{figure*}
\begin{tabular}{llll}
(a) & \includegraphics[width=1\columnwidth]{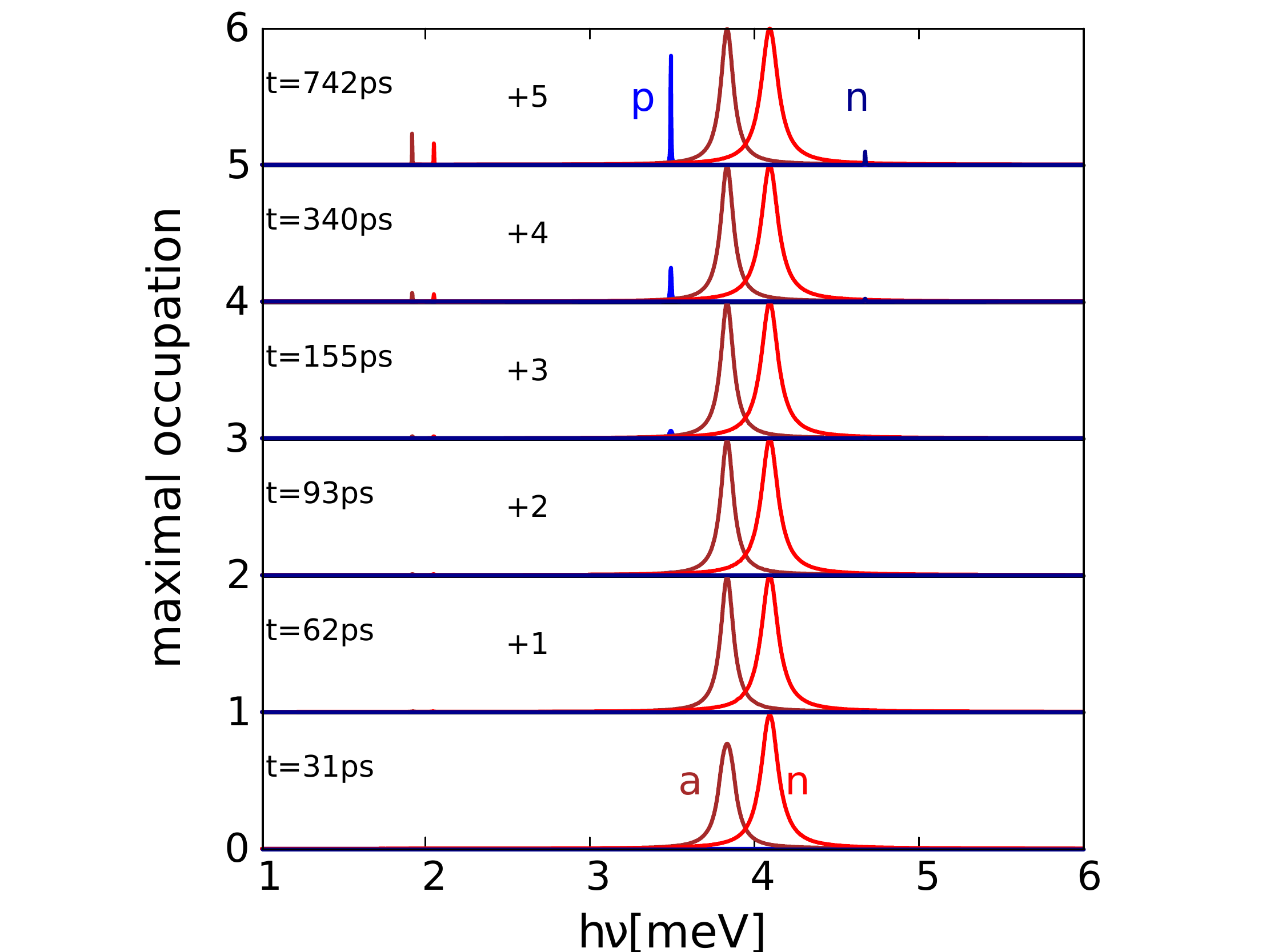} &
(b) & \includegraphics[width=1\columnwidth]{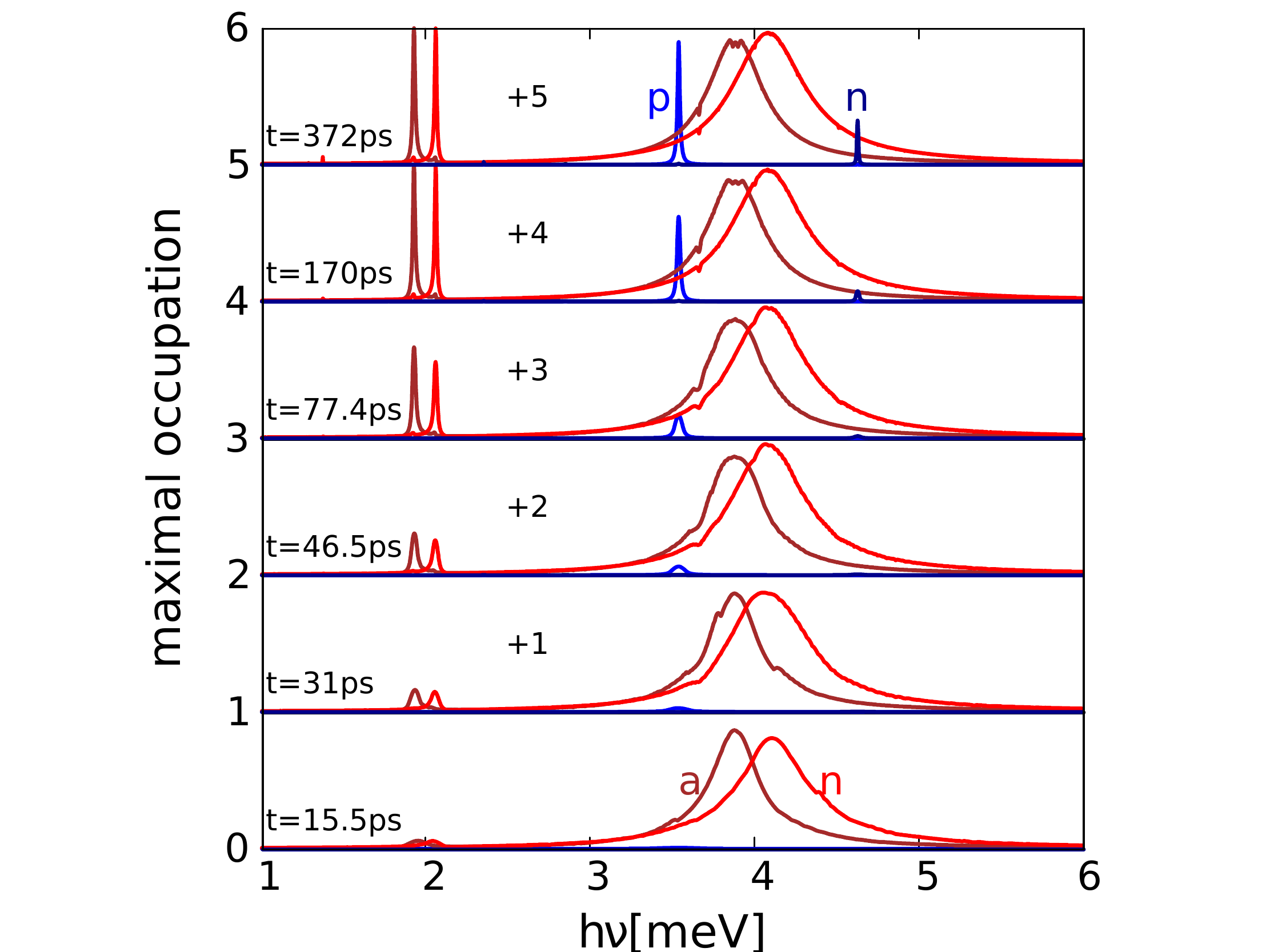} \\ 
\end{tabular}
  \caption{ The results of the time-dependent simulation for the AC in-plane electric field of the amplitude $F_{ac}=100$ V/cm (a) and $F_{ac}=400$ V/cm (b) for the vertical electric field of $F_z=0.5$V/\AA\  
and the vertical magnetic field of $B=1$ T. The electron is set initially in the p$\downarrow$ ground state and AC electric field  $ex F_{ac}\sin(2\pi h\nu t)$ is applied.  The maximal occupation of the stationary eigenstates is given for the duration of the simulation  increasing 
for the higher curves. The results are shifted by +1 for subsequent plots and the duration times are listed in the figure. The dark red curve labeled by a and the lighter red curve labeled by n indicate spin conserving transitions 
p$\downarrow\rightarrow a \downarrow$ and p$\downarrow\rightarrow n \downarrow$, respectively. The light blue curve labeled by p and the dark blue curve labeled by n correspond to spin-flipping transitions  p$\downarrow\rightarrow p \uparrow$ and p$\downarrow\rightarrow n \uparrow$, respectively.
 }
 \label{trak5}
\end{figure*}

\begin{figure*}
\begin{tabular}{llll}
(a) & \includegraphics[width=.8\columnwidth]{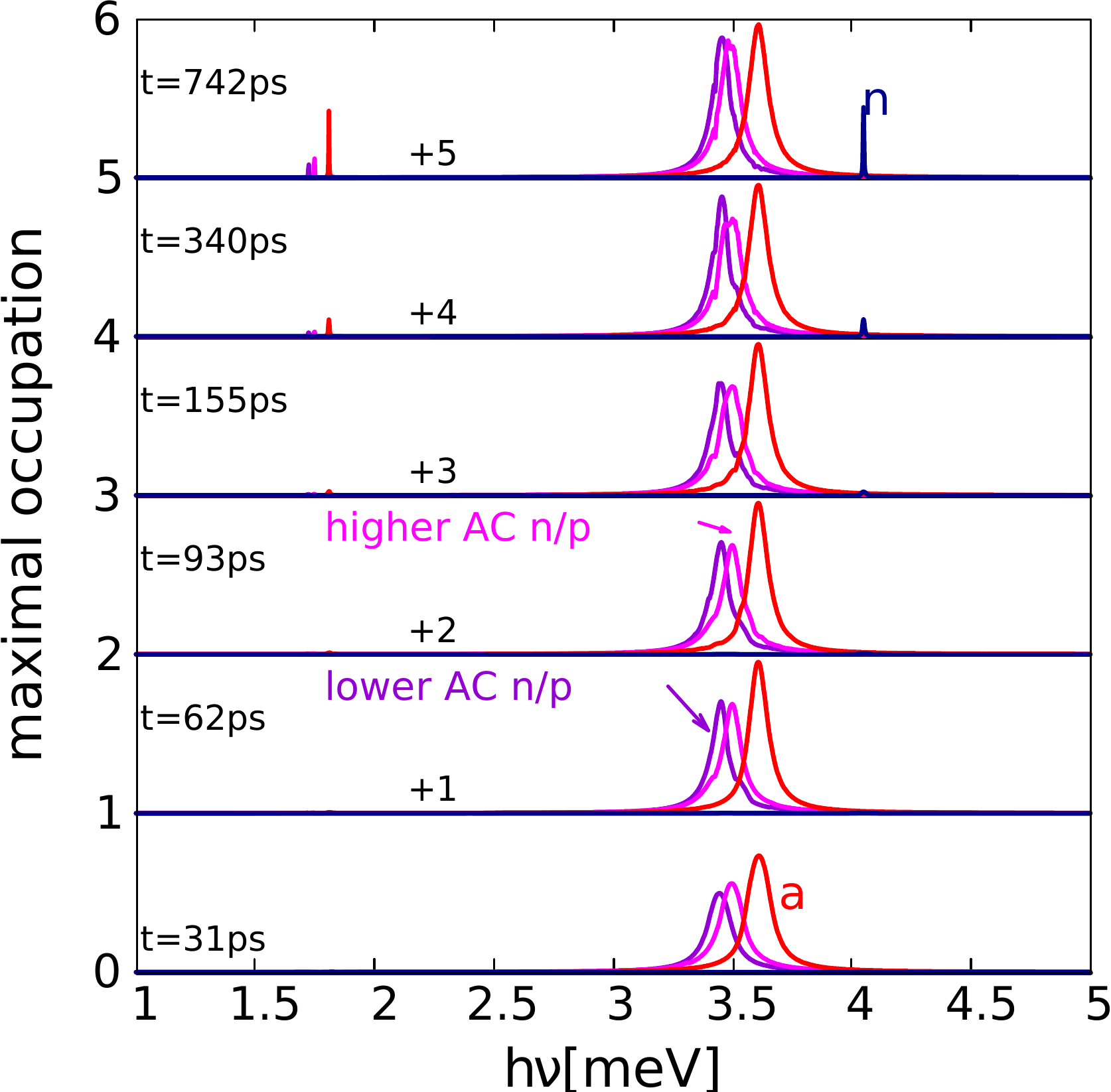} &
(b) & \includegraphics[width=.8\columnwidth]{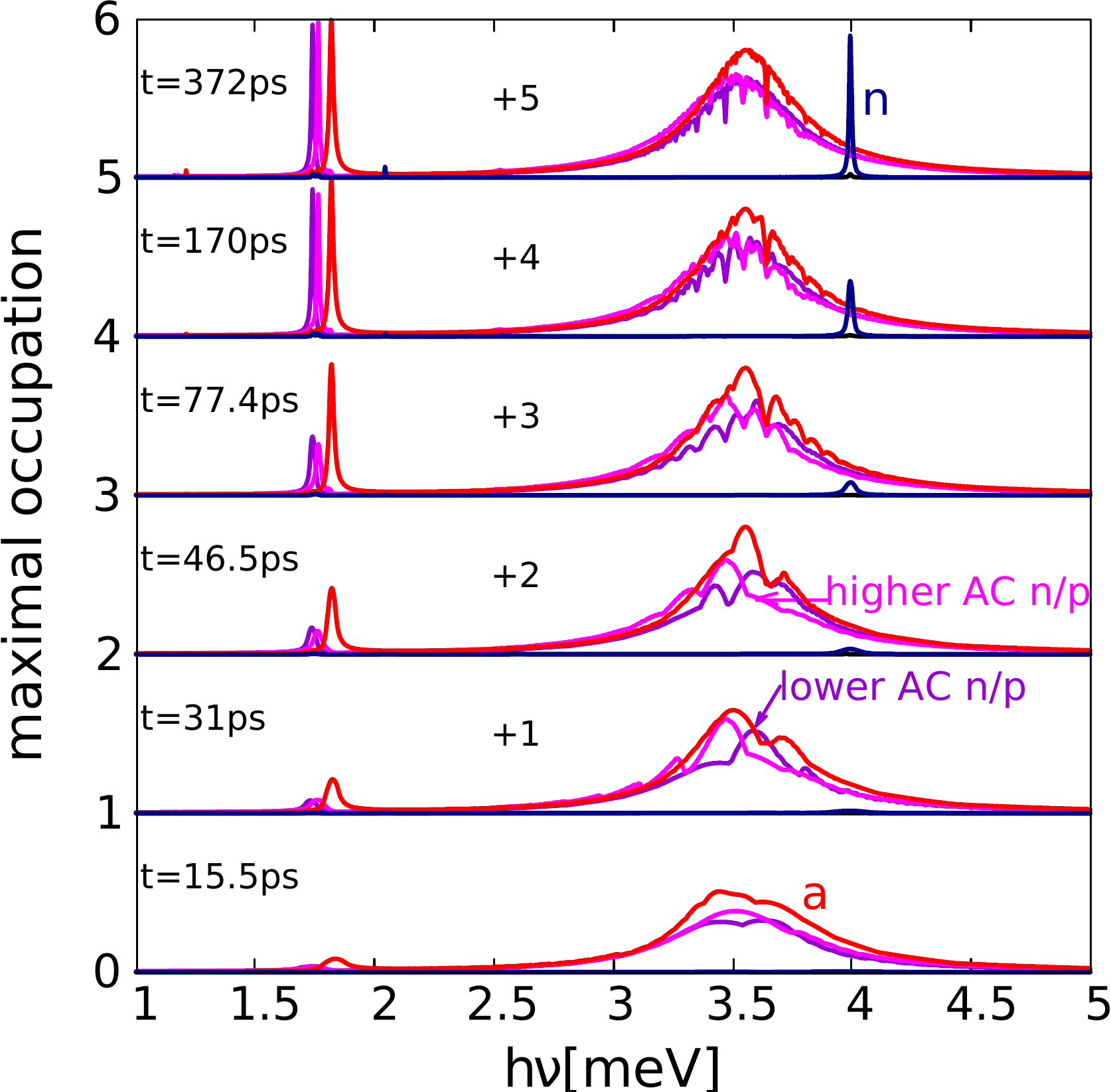} \\ 
(c) & \includegraphics[width=.8\columnwidth]{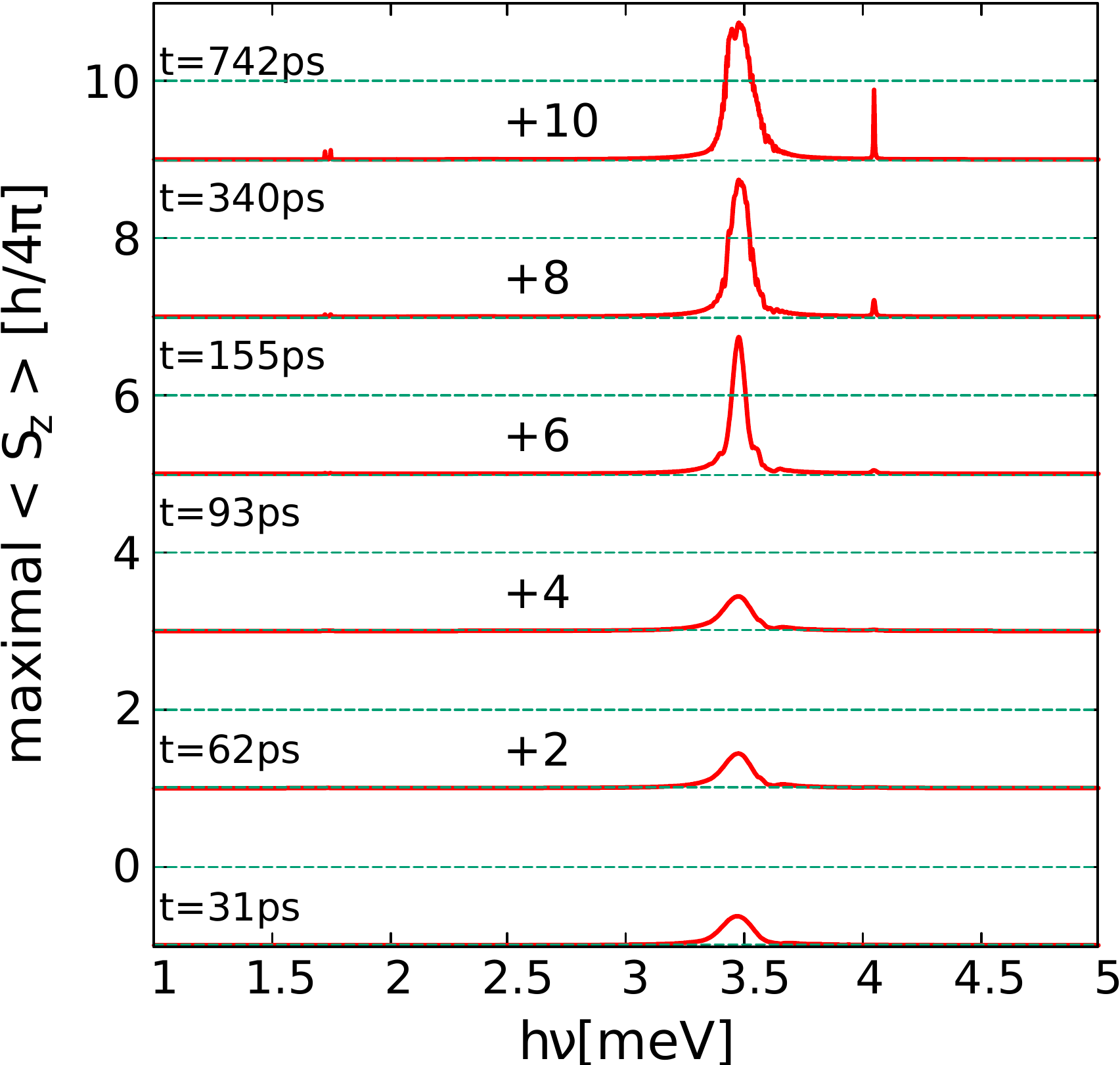}  & (d) & \includegraphics[width=.8\columnwidth]{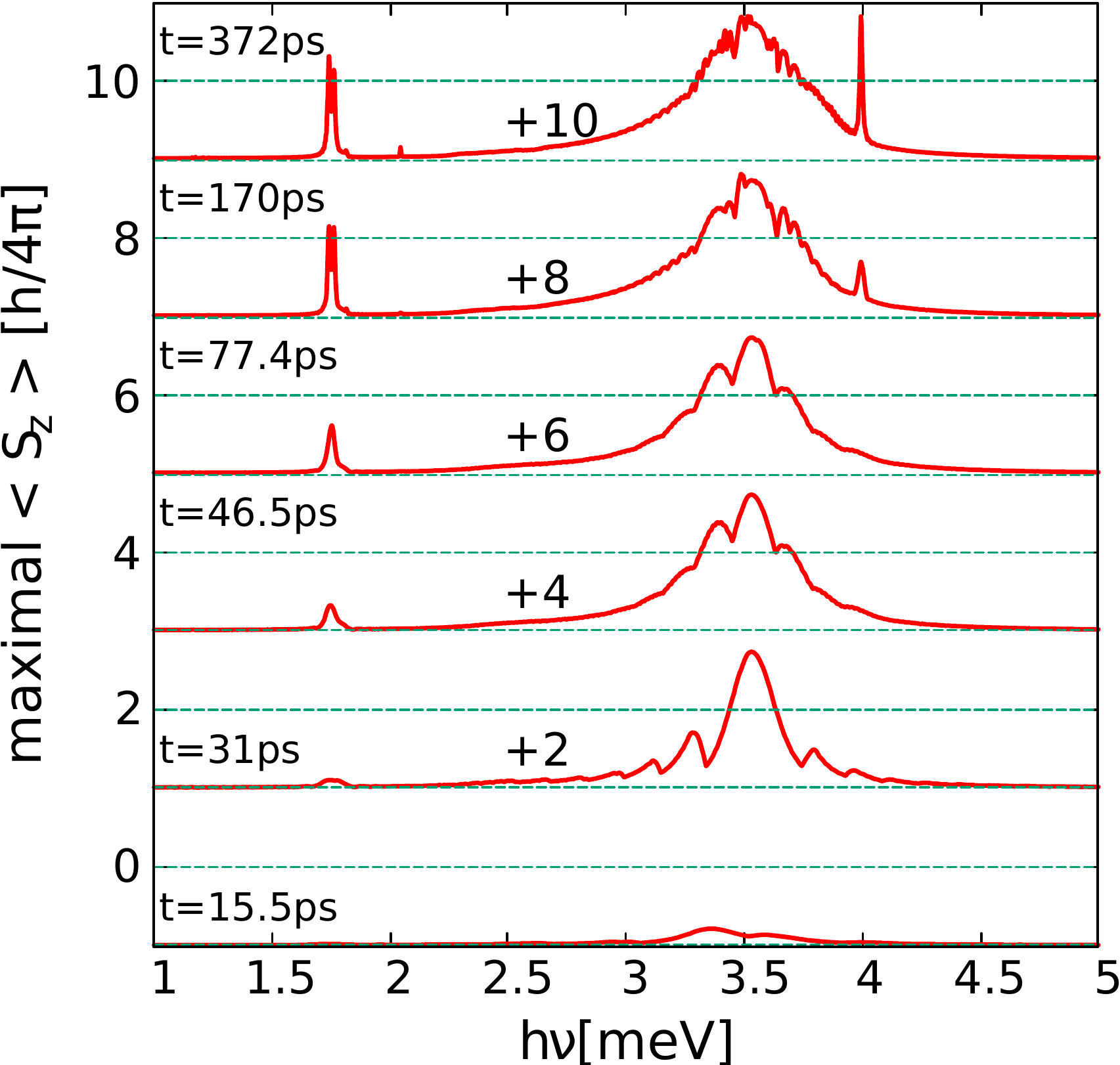} 
\end{tabular}
    \caption{
(a,b) Same as Fig. \ref{trak5} only for $F_z=0.679$V/\AA. The left panels (a,c) were plotted for the amplitude of the AC electric field $F_{ac}=100$ V/cm and the right ones (b,d) for $F_{ac}=400$ V/cm.  The pink and 
the violet curves indicate the transitions to states that
participate in the p$\uparrow\leftrightarrow n\downarrow$ avoided crossing [Fig. \ref{wpulu}(c)], respectively.
The blue curve labeled by n corresponds to transition to n$\uparrow$ state and the red curve labeled by a to transition
to a$\downarrow$ state. 
(c,d) Indicate the maximal projection of the spin for p$\downarrow$ state in the initial condition. 
 }
 \label{trak679}
\end{figure*}

On the scale of Fig. \ref{wpulu} the contribution of the Rashba terms cannot be resolved. The energy levels near the avoided crossing in Fig. \ref{wpulu}(c) are determined mostly by the extrinsic Rashba interaction ($t_3$).
For the intrinsic Rashba interaction excluded ($t_1=0$) the spectrum of Fig. \ref{wpulu}(c) does not change at the scale of the figure.
Both Rashba interactions open  avoided crossings between the same pairs of states.
For the  width of the avoided crossing defined as the spacing between the anticrossing energy levels
in the middle of the crossing at the scale of $F_z$, the widths of the avoided crossing
in Fig. \ref{wpulu}(c) are  45 $\mu$eV for p$\uparrow\leftrightarrow$n$\downarrow$ and 31 $\mu$eV
for a$\downarrow\leftrightarrow$n$\uparrow$. When the external Rashba term is switched off ($t_3=0$),
the corresponding widths narrow down to 5$\mu$eV and 3$\mu$eV, respectively.

In Figure \ref{wpulu}(b) with the dotted lines we plotted the matrix elements as obtained for $t_3=0$. 
The locally maximal values of the matrix elements remain the same, only the width of the maxima is reduced. 
Therefore, a larger precision would be required  in order to tune into the avoided crossing with the vertical electric field $F_z$. For the external Rashba interaction excluded ($t_3=0$) the matrix elements return
to small values at large $F_z$ above the avoided crossings. 

For the value of $t_3$ kept unchanged but $t_1$ set to zero the results cannot be distinguished
from the ones plotted with the solid lines in Figure \ref{wpulu}(b).

\subsection{The spin resonance}

For simulation of the electric dipole spin resonance the external magnetic field of 1T was set to lift the degeneracies.
 In the initial condition the electron was set in the p$\downarrow$ ground-state.
For $t>0$ the AC electric field of the amplitude $F_{ac}$ and frequency $\nu$ is applied [Eq. (4)].
For each value of $\nu$ we monitored in time the maximal occupation of the excited states of the 
stationary Hamiltonian by projection of the wave function on the $H_0$ eigenstates basis. 

We studied the EDSR  for $F_z=0.5$ V/\AA\; [Fig. \ref{trak5}] i.e. in the neighborhood of the n$\downarrow\leftrightarrow$p$\uparrow$ avoided crossing  (see Fig. \ref{wpulu}),
and in the center of this avoided crossing for $F_z=0.679$ V/\AA\; [Fig. \ref{trak679}]. For $F_z=0.5$ V/\AA\; the spin of the eigenstates is polarized parallel or antiparallel to the $z$ axis.
Within the avoided crossing  $F_z=0.679$ V/\AA\; [Fig. \ref{trak679}] the spins are no longer polarized.

The results for $F_z=0.5$ V/\AA\; in Fig. \ref{trak5} show wide maxima corresponding to 
spin-conserving transitions to a$\downarrow$ and n$\downarrow$ states.
For $F_{ac}=100$ V/cm,  the maximal occupation probabilities of the excited $\downarrow$ states reach 1  for AC pulse duration of several dozens of ps [Fig. \ref{trak5}(a)] provided that the AC frequency matches the resonance. The spin-flip transitions are resolved  later in the simulation and the width of the resonant peaks is  smaller. The transition to p$\uparrow$ state for $F_z=0.5$ V/\AA\; is distinctly faster than the one to n$\uparrow$. For $F_z=0.5$ V/\AA\; the avoided crossing involving p$\uparrow$ is closer [cf. Fig. \ref{wpulu}(b)]. In the  conditions  of Fig. \ref{trak5}(a) the spin flips occur according to the two-level Rabi resonance.

Figure \ref{trak5}(b) shows the results for the amplitude of the AC field increased to $400$ V/cm. 
The transitions rates to the considered excited states are increased. However, the spin-conserving transitions to a$\downarrow$ 
and n$\downarrow$ become wide and overlap on the $\nu$ scale with each other and with the spin flipping transitions. 
In Fig. \ref{trak5}(b) the two narrow orange and brown peaks near $h\nu\simeq 2$ meV are the two-photon transitions to a$\downarrow$ and n$\downarrow$ states at half the resonant single-photon frequency.

In the center of the avoided crossing [Fig. \ref{trak679}] the transitions involving the spin inversion
 appear now nearly as fast as the spin-conserving transition to a$\downarrow$ state which is higher in the energy. 
A wide peak is observed at the energy of the avoided crossing [cf. Fig. \ref{trak679}(a,c) and Fig. \ref{trak679}(b,d)] and a sharp one at the higher energy for the transition to n$\uparrow$ state off the avoided crossing. 

The two-photon spin-flipping transitions are observed for the larger AC amplitude [Fig. \ref{trak679}(d)]. Since the time evolution within the avoided crossing involves more than just two states -- the transition does not have the typical Rabi dynamics. None of the states within the avoided crossings has a well defined spin orientation. In consequence, the spin flip probability, which acquires large values already after $\simeq 30$ ps does not reach 100\% even for the longest times studied. Nevertheless, the present results indicate that one can tune the  external static fields to conditions in which the Rashba interactions -- notoriously weak in silicene \cite{Ezawa} -- can be harnessed for fast spin transitions. These transitions  should produce a strong signal in the EDSR spectra  by 
effective lifting of the Pauli blockade of the current \cite{edsr1,edsr2,edsr3,edsr4,edsr5,edsr6,edsr7,extreme,stroer,edsrcnt1,edsrcnt2,edsrcnt3}. The fast spin flips run by electric pulses should  be useful
for experimental studies of the spin structure and interactions. 
A workpoint for the two-level Rabi dynamics outside the avoided crossings can also be selected.

The results of this work are obtained for the side length of the hexagonal flake of $s=18.64$ nm.
We checked how the size of the hexagonal flake influences the conditions and the dynamics of the EDSR. For that purpose we considered a smaller flake with side length $s_s=9$ nm and a larger one with $s_l=24$ nm.
The spin-orbit coupling splitting at $B=0$ is $\Delta=3.09$ meV, 3.05 meV and 3.03 meV
for $s_s$, $s$ and $s_l$ respectively. The magnetic field for which an avoided crossing 
of n$\uparrow$ -- a$\downarrow$ energy levels strongly depends on the size of the flake,
and equals 23.7 T, 8.73 T and 3.25 T, for $s_s$, $s$ and $s_l$ respectively. 
The width of the avoided crossing -- involving spin mixing between the states -- is smaller for smaller flakes. We attribute this fact to  the
larger Zeeman interaction for higher $B$. The corresponding widths are 23$\mu$eV, 28.3$\mu$eV and 46 $\mu$eV, in the same order as above. The spin-flip time for the smallest flake with $s_s$ in the center of the n$\uparrow$ -- a$\downarrow$ avoided crossing for $F_{ac}=400$ V/cm
is 23 ps, of the order found in Fig. \ref{trak679}(b,d).

\subsection{Circular confinement and removal of the edge effects}

The spin flips demonstrated above were accelerated within the range of avoided crossings [cf. Fig. \ref{fk25}(b)] involving the  n states which at $B=0$ appear close above  the ground-state.  In the absence of the spin-orbit interaction and without the intervalley scattering all the energy levels
at $B=0$ should be fourfold degenerate: with respect to the valley and the spin. 
At $B=0$ the ground-state in the hexagonal armchair flake in Fig. \ref{fz0}(a) is indeed fourfold degenerate -- with respect to the current orientation and to the spin, but the n states
are only twofold degenerate -- with respect to the spin only.
The twofold degenerate levels at $B=0$ can only result from
the intervalley scattering which in the studied flake is induced by the armchair edge.
In order to estimate the effects of the intervalley scattering  it is instructive
to decouple the localized states from the edge. 

Moreover, the confined states and the spin-orbit coupling mechanism could be more precisely described in circular confinement
using  the low-energy continuum approximation \cite{gru,nori}. In this case the angular momentum can be used to characterize the states provided that the intervalley scattering is removed. 
The reason for the latter is the following. In presence of the intervalley coupling 
the wave function in the  sublattices can be described   by 
 \begin{equation}
\psi_A({\bf r})=\exp(i{\bf K}\cdot {\bf r})\phi_A({\bf r})+\exp(i{\bf K'}\cdot {\bf r})\phi_{A'}({\bf r}), \label{mix} \end{equation}
and
 \begin{equation}
\psi_B({\bf r})=\exp(i{\bf K}\cdot {\bf r})\phi_B({\bf r})+\exp(i{\bf K'}\cdot {\bf r})\phi_{B'}({\bf r}), \label{mix} \end{equation}
 where $\phi_A,\phi_B$  are the envelope functions corresponding to the ${\bf K}$ valley,
and  $\phi_{A'},\phi_{B'}$ to the ${\bf K'}$ valley.
The armchair edge does not support localized states, hence $\psi_A({\bf r})=0,\psi_B({\bf r})=0$ at the edge, which implies the boundary conditions on the envelope functions of the form \cite{zarenia} \begin{equation} 
\phi_A({\bf r})=-\exp(i({\bf K'-K})\cdot {\bf r})\phi_{A'}({\bf r}), \label{bca}\end{equation} and
\begin{equation}\phi_B({\bf r})=-\exp(i({\bf K'-K})\cdot {\bf r})\phi_{B'}({\bf r}). \label{bca2}\end{equation}
Since  the  intervalley distance $|{\bf K'-K}$| is large, the exponent in Eq. (\ref{bca}) oscillates rapidly at the atomic scale
and lifts the isotropy of the problem even for a circular edge of the flake.

In this section we introduce 
a circular confinement and decouple the confined states from the edge. 
For that purpose we introduce an additional gap modulation \cite{nori} within the flake.
Namely, we consider a modified Hamiltonian 
\begin{equation}
H_W=H_0 +  \sum_{k,\alpha} W ({\bf r_k}) c^\dagger_{k\alpha}c_{k\alpha}.  \label{kas} 
\end{equation}
where $H_0$ is defined by Eq. (\ref{haha}). In Eq. (\ref{kas}), $W({\bf r_k})=W_0 \; \mathrm{sign}(z_k)$ for $r_k\geq R$ and $W({\bf r_k})=0$ for $r_k<R$. The extra term with $W$  widens the energy gap for $r>R$ \cite{nori} which results in the carrier confinement for energies close to the Dirac point.

\begin{figure}
 \includegraphics[width=1.2\columnwidth]{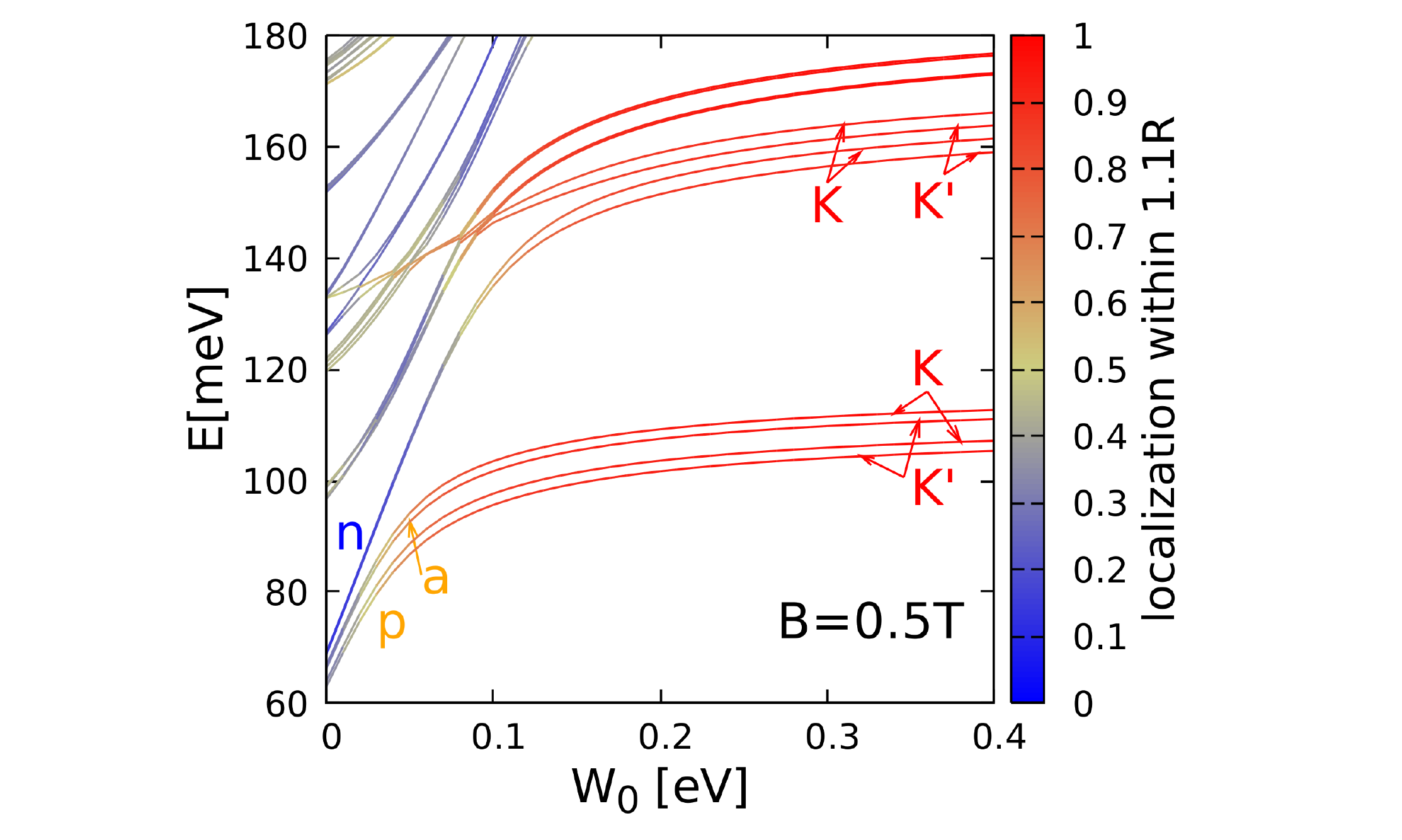} 
    \caption{The energy spectrum for Hamiltonian \ref{kas} as a function of the $W_0$ parameter that opens the energy gap for $r>R=10$ nm.
The color of the lines indicate the electron localization within the radius $1.1R$ from the origin. 
The results are calculated for $F_z=0.25$ V/\AA\; with spin-orbit interactions and $B=0.5$ T. 
The lowest energy levels at $W_0$ are denoted by $n,p$ and a as above in this work. For larger $W_0$ one can attribute valley index $K,K'$
to the levels.}
 \label{cbin}
\end{figure}

The resulting confinement is illustrated in Fig. \ref{cbin} for $R=10$ nm. The energy levels that are plotted as functions of  $W_0$
turn red when the state gets localized within $r<1.1R$. In Fig. \ref{cbin} all the spin interactions are present and $B=0.5$ T is applied in order
to split the degeneracies. We can see that the pair of n energy levels joins two other energy levels at $W_0>\sim 0.2$ eV and the resulting quadruple
moves parallel for larger $W_0$. The appearance of the quadruple is a signature of lifting the intervalley scattering by separation of the states
from the edge. 
The quadruple is formed by two doublets which at $W_0=0$ are separated by a large energy spacing of about 70 meV. Only due to the strong intervalley coupling the n energy levels appear low in the energy spectrum close to the ground-state at $W_0=0$.
The chance for accelerated spin flips at relatively low $B$ -- where the avoided crossings appear -- results from the intervalley coupling.

\label{dirac}

\begin{figure}
\begin{tabular}{l}
a) \includegraphics[width=.8\columnwidth]{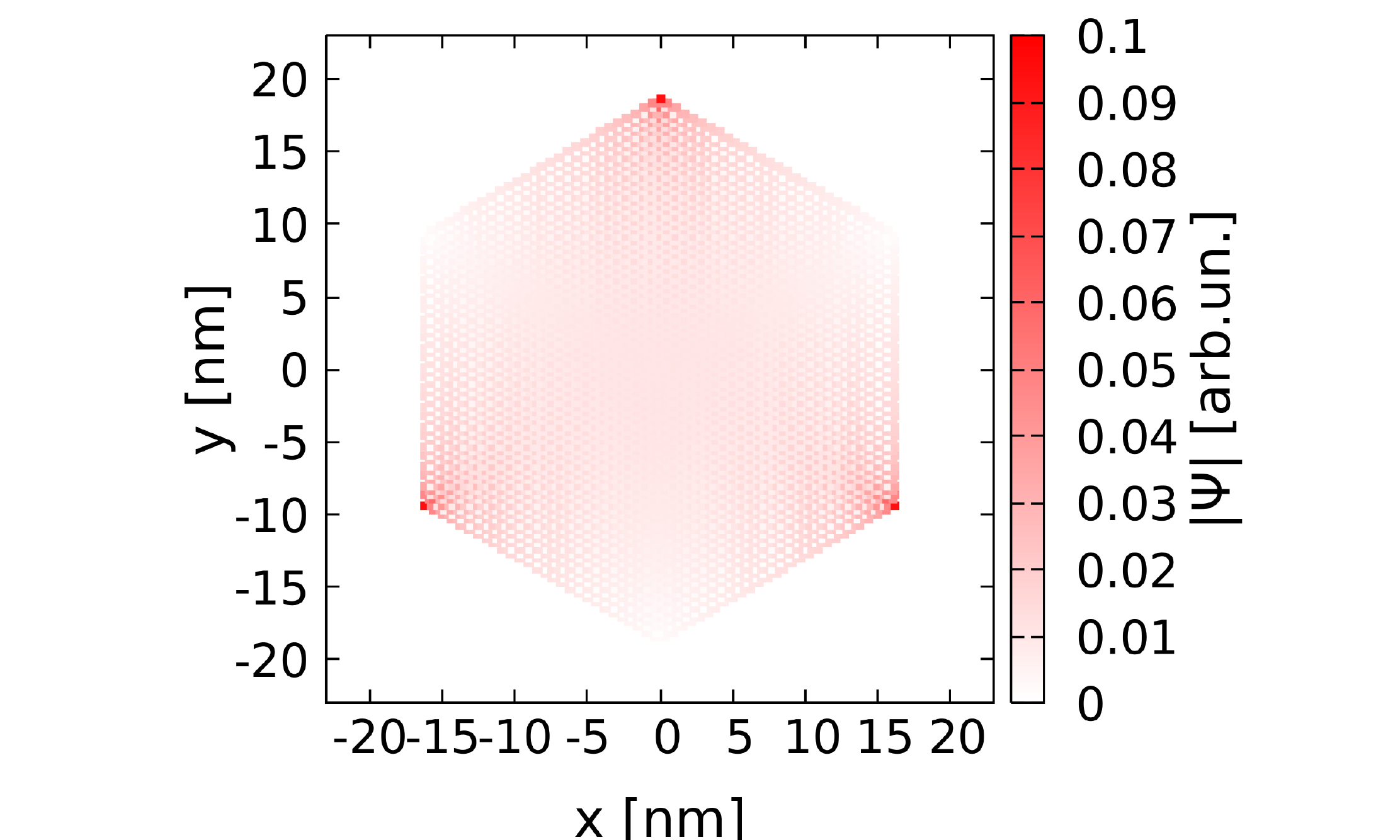}\\
b) \includegraphics[width=.8\columnwidth]{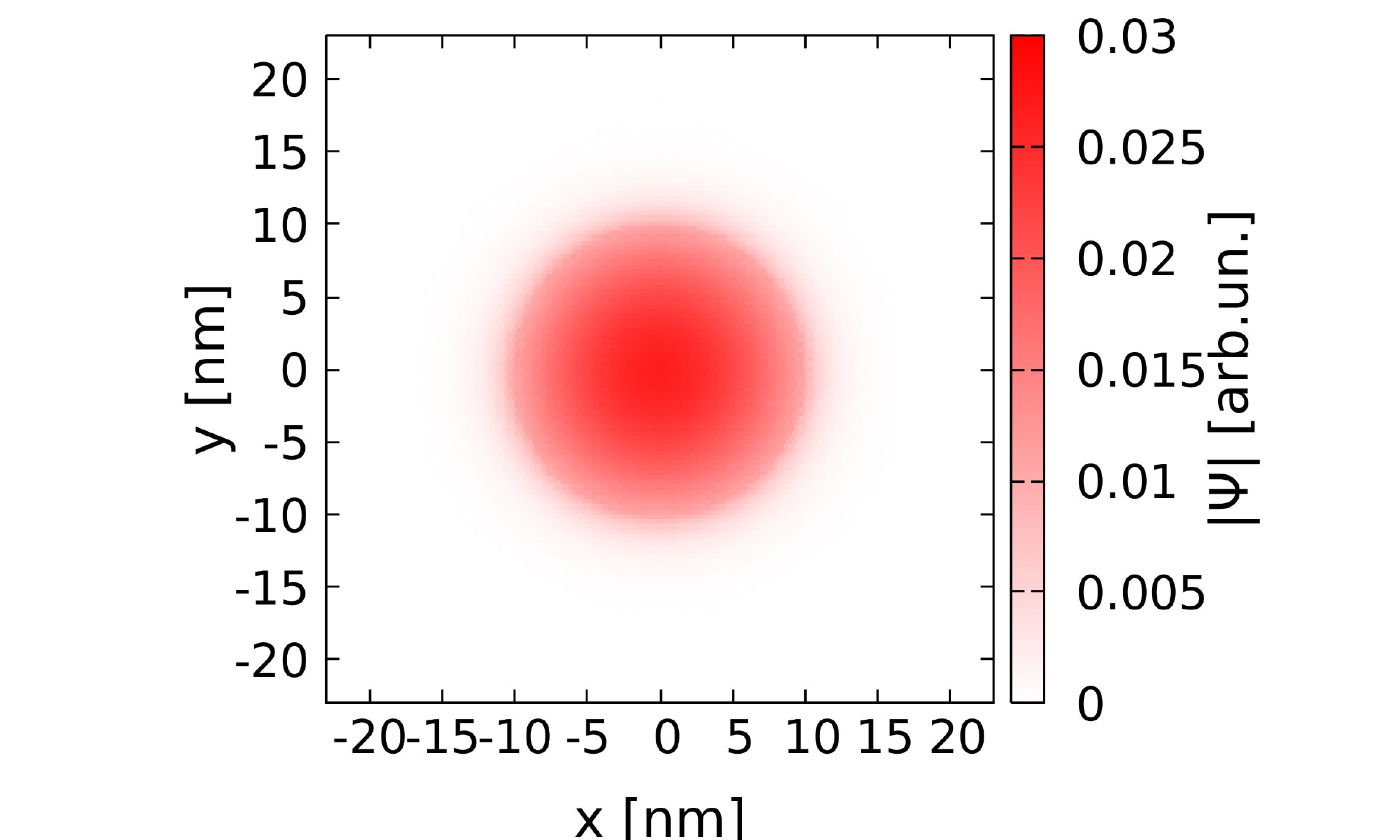}  \end{tabular}
    \caption{Absolute value of the wave function  at the $A$ sublattice in the conduction band ground state at $W_0=0$ (a) and $W_0=0.3$ eV (b) for $B=1$ T.
Other parameters as in Fig. \ref{cbin}.}
 \label{fuks}
\end{figure}
 
The absolute value of the ground-state wave function on the $A$ lattice is displayed  for $W_0=0$ 
in Fig. \ref{fuks}(a) 
and for $W_0=0.3$ eV in Fig. \ref{fuks}(b).
Figure \ref{fuks}(a) contains a checkerboard pattern near the corners of the hexagon and the results of Fig. \ref{fuks}(b) are smooth.
The checkerboard results from the rapid wave function oscillations (see Eq. (\ref{mix})) that follows the intervalley scattering induced by the edge. Even for smooth $|\phi_A|$ and $|\phi_{A'}|$,
the exponent of Eq. (\ref{mix}) generates rapid variation of the absolute value of $\psi_A$.
The variation  is removed once the coupling of the localized state to the edge is lifted (Fig. \ref{fuks}(b)).

The quantum numbers for the localized states  can be conveniently explained in the context of the magnetic field dependence of the energy spectra.
For identification of the tight-binding states they are compared to the ones obtained with the continuum approximation \cite{Ezawa12a},
for which we keep track of the intrinsic spin-orbit interaction and neglect the Rashba terms. 
For the vector wave function $\Psi=(\Psi_A,\Psi_B)^T$
the continuum Hamiltonian reads \cite{Ezawa12a}
\begin{equation} H_\eta = \hbar V_F \left(k_x \tau_x -\eta k_y \tau_y \right)+U({\bf r})\tau_z+\frac{g\mu_B B}{2}\sigma_z -\eta \tau_z \sigma_z 3\sqrt{3} t_2, \label{cont}\end{equation}
with $U({\bf r})=eF_z z+W({\bf r})$, 
$\eta=1$ for $K$ valley and $\eta=-1$  for $K'$ valley, ${\bf k}=-i\nabla+\frac{e}{\hbar} \vec{A}$, and $V_F=\frac{{3at}}{2\hbar}$ is the Fermi velocity, with the nearest neighbor distance $a=2.25$\AA. $\tau_x$, $\tau_y$ and $\tau_z$ are the Pauli
matrices in the space spanned by the sublattices. 

With the  symmetric gauge $\vec{A}=(-By/2,Bx/2,0)$ the Hamiltonian eigenstates $\Psi_\eta$ can be characterized by eigenvalues of the orbital momentum operator $j_z=l_z {\bf I} +\eta \frac{\hbar}{2} \tau_z$, where $l_z=-i\hbar \frac{\partial }{\partial \phi}$ 
and ${\bf I}$ is the unit matrix. The  eigenfunction is then $\Psi_\eta=\left[ f_A (r) \exp(im\phi) ,f_B (r) \exp(i(m+\eta)\phi) \right]^T$
where $m$ is an integer.
Summarizing, the continuum Hamiltonian (\ref{cont}) eigenstates have a definite $z$ component of the spin, the valley index, and the angular momentum. 
We label the Hamiltonian eigenstates with quantum number $j=m +\eta/2$.

\begin{figure}
\begin{tabular}{ll}
a) \includegraphics[width=.4\columnwidth]{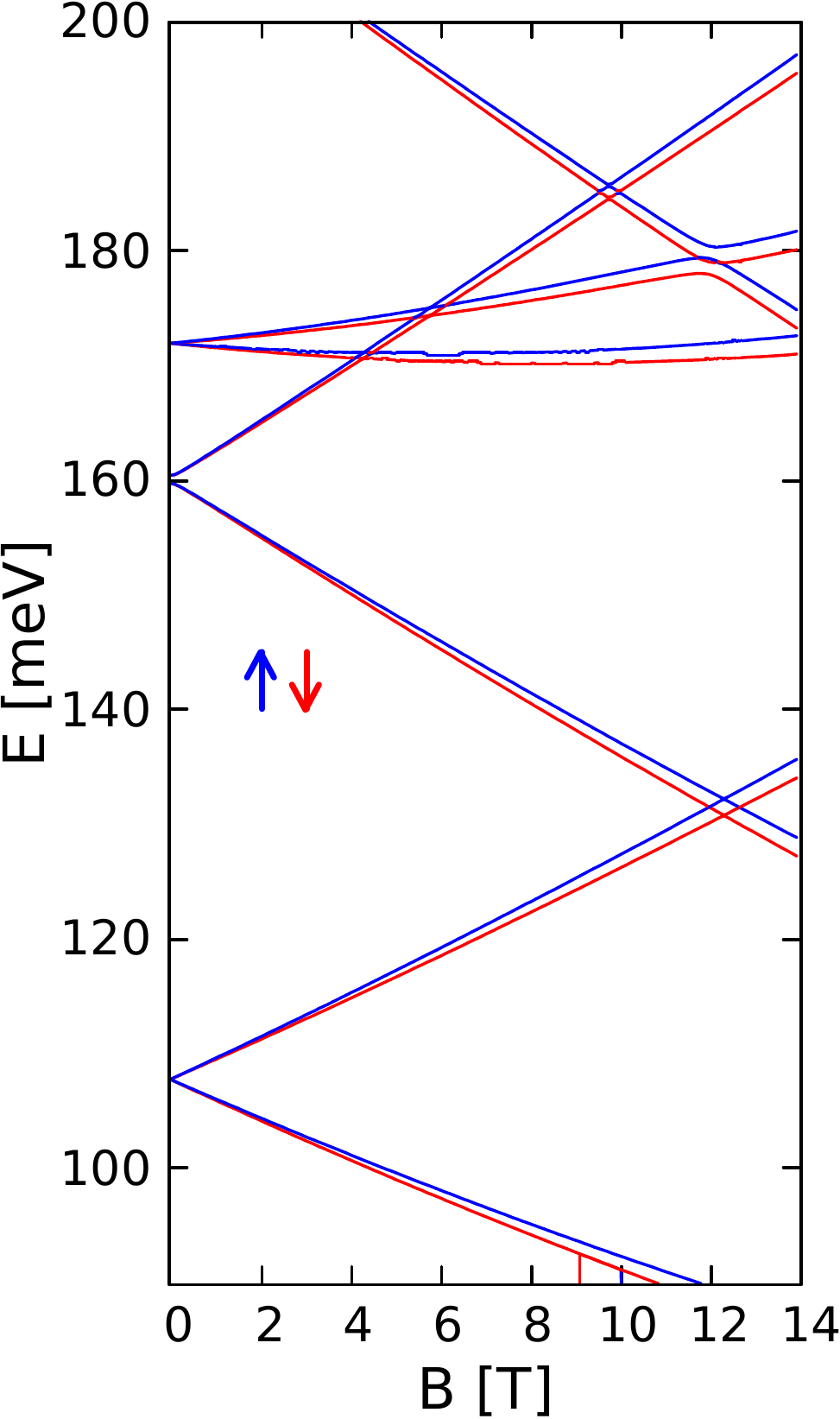} & b) \includegraphics[width=.48\columnwidth]{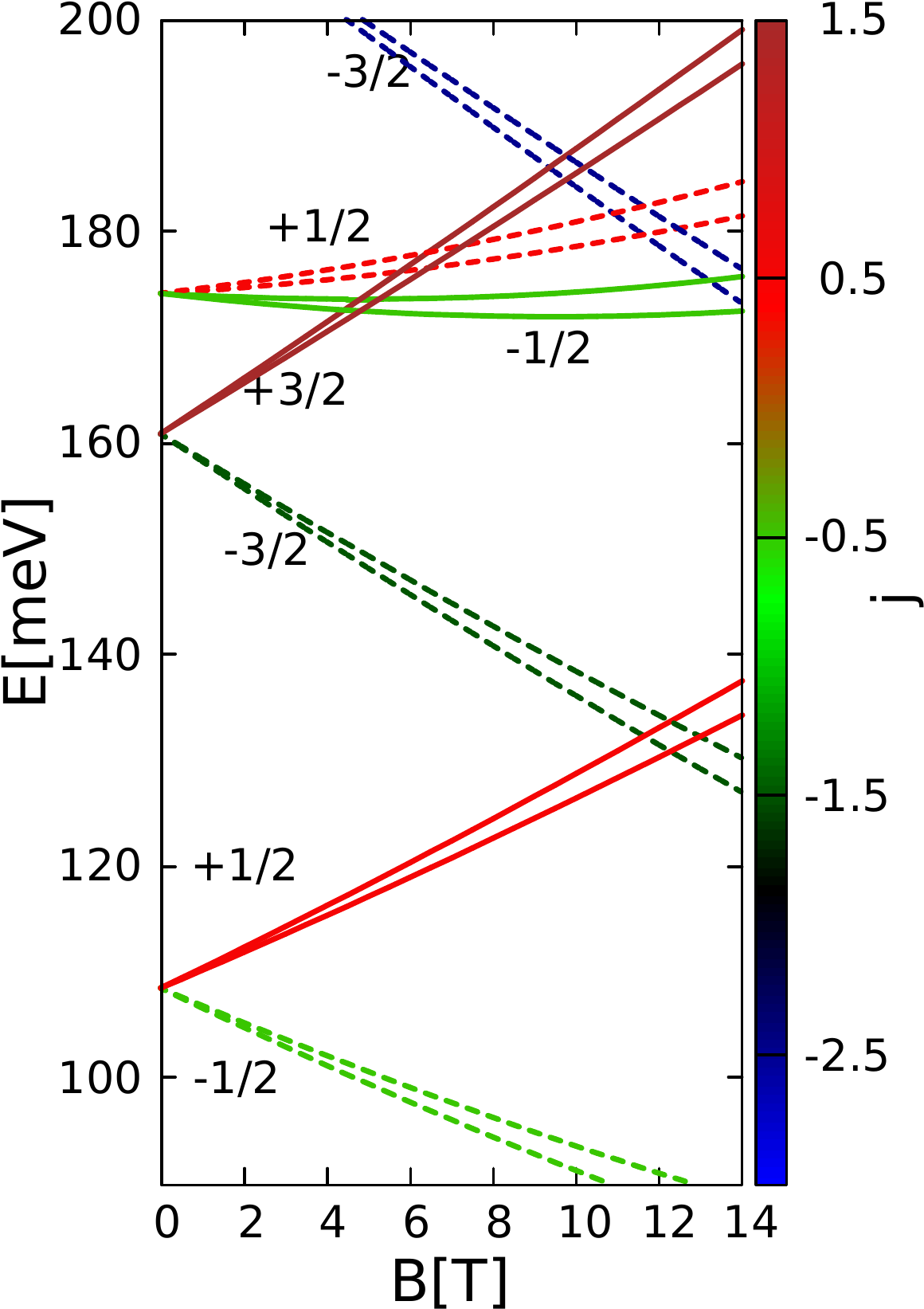}\\
c) \includegraphics[width=.4\columnwidth]{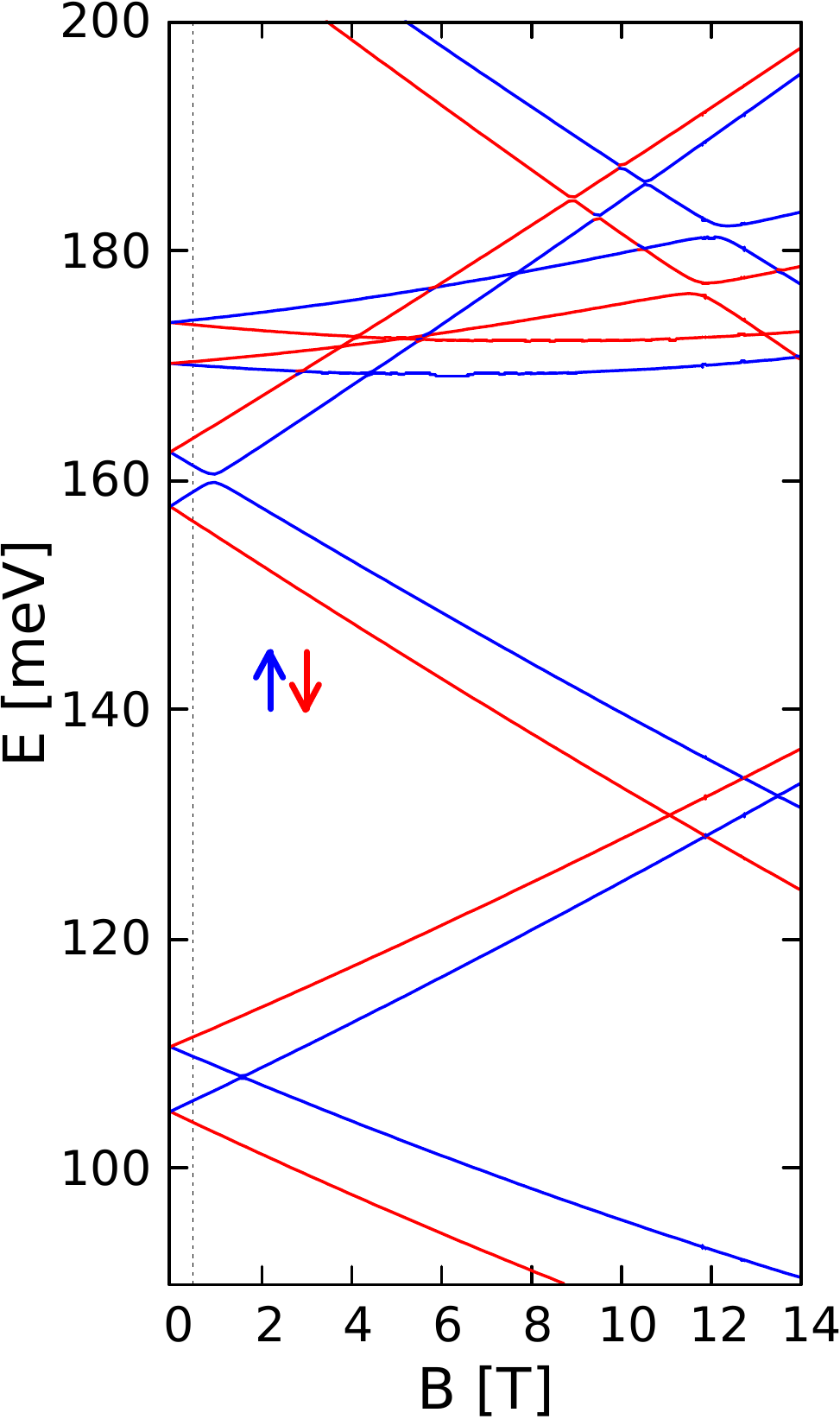} & d) \includegraphics[width=.48\columnwidth]{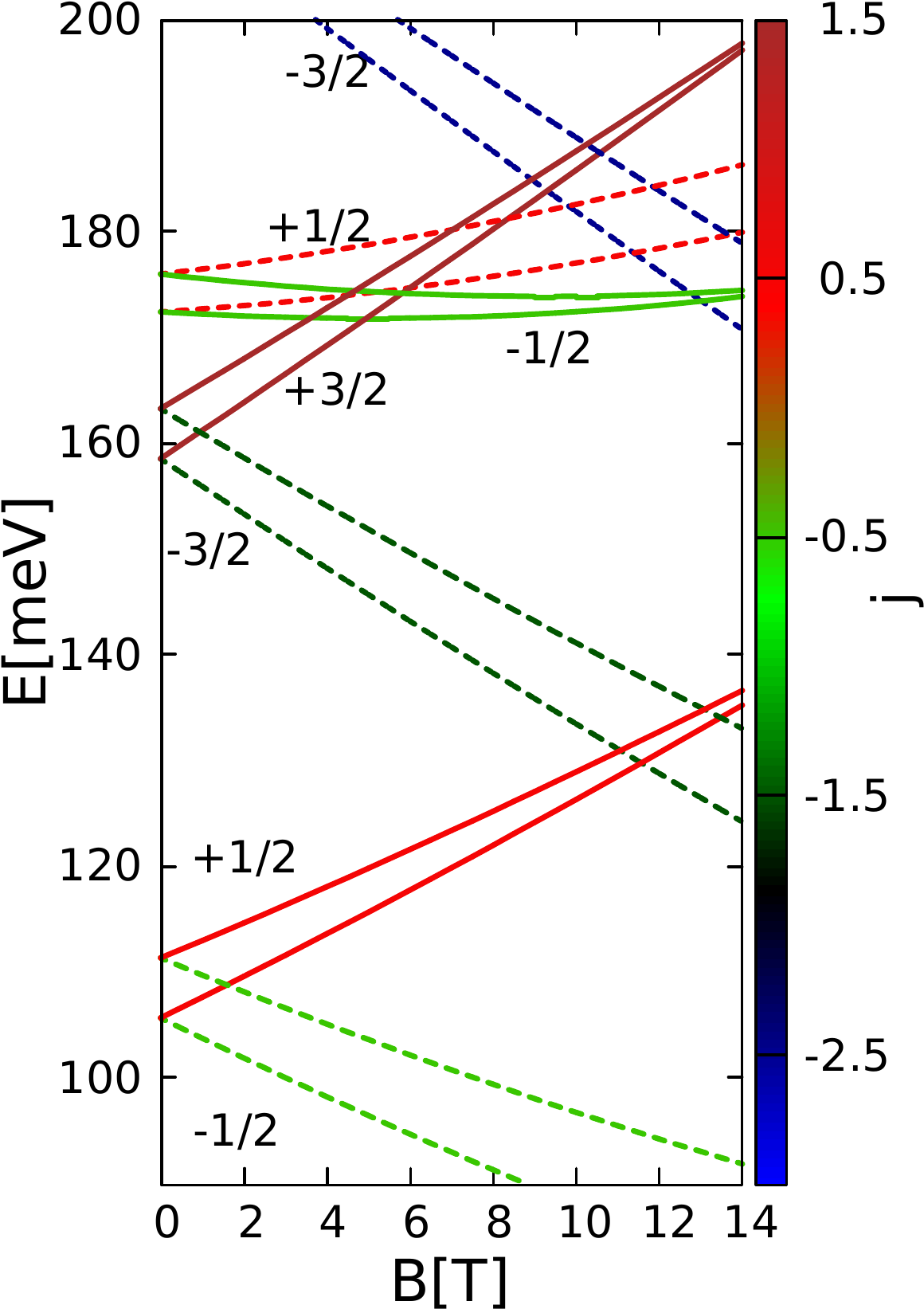} \end{tabular}
    \caption{Energy spectra as obtained with the atomistic tight-binding (a,c) and
the continuum approach (b,d) for $F_z=0.25$ V/\AA\; and $W_0=0.3$ eV.
The spin-orbit interactions are absent in (a,b) and present in (c,d). 
In (a,c) the red and blue lines correspond to spin-down and spin-up states.
In (b,d) the dashed lines stand for the $K'$ valley and the solid lines
to the $K$ valley.  In (b,d) the color of the lines and the fractions near the curves indicate the orbital angular momentum $j$.
The vertical line in (c) shows $B=0.5$ T, for which Fig. \ref{cbin} was calculated.
}
 \label{mama}
\end{figure}

Figure \ref{mama}(a,c) shows the energy spectrum as obtained with the tight-binding approach
used above with [Fig. \ref{mama}(a)] or without [Fig. \ref{mama}(c)] the spin-orbit coupling for $W_0=0.3$ eV.
The results of a continuum approach are displayed in Fig. \ref{mama}(b) without and in Fig. \ref{mama}(b)  with
the spin-orbit interaction. The dashed (solid) lines indicate the $K'$ ($K$) valley levels. 
The results for both approaches are nearly identical and Fig. \ref{mama} contains all
the information on the corresponding energy levels with respect to both angular momentum, the valley, and the spin.   
In these conditions one can indicate the mechanism of the spin-orbit coupling more precisely: the spin-orbit interactions splits the fourfold degenerate states of Fig. \ref{mama}(a,b)
to pairs of doublets in the following manner: at $B=0$ in the lower doublet we find 
$K'\downarrow$ and $K\uparrow$ states, and in the higher doublet $K\downarrow$ and $K'\uparrow$ states.
 This form of the spin-orbit coupling splitting is observed in carbon nanotubes \cite{kuem}.

In Fig. \ref{mama} the $K'$ $j=-3/2$ energy levels  cross the $K$ $j=+1/2$ energy levels near 12 T. These
crossings are counterparts of the avoided crossings between a,p and  n energy levels 
for the hexagonal flake that increased the spin flip transition matrix element [cf. Fig. 5].
Now, in the tight-binding calculations [Fig. \ref{mama}(c)] that accounts for the Rashba interaction -- we do not resolve any repulsion of the energy levels,
or in any case the width of the avoided crossing is smaller than 0.1 $\mu$eV. The Rashba spin-orbit interaction -- that is accounted for in Fig. \ref{mama}(c) does not open couple energy levels associated to opposite valleys.
 As far as the transition matrix elements from the ground-state to five lowest-energy excited states in Fig. \ref{mama}(c) are concerned
 -- the only large one  is from $K',j=-1/2 \downarrow$ to $K',j=-3/2 \downarrow$ and equals 3.35 nm at $B=0$. The $x$ matrix elements for the spin-flipping transitions
as calculated with the atomistic tight binding do not exceed $0.004$ nm.

 Based on the result of Fig. \ref{cbin} and Fig. \ref{mama}(d) we can see that the n states for the hexagonal flake were mixtures of $K$ and $K'$ states with $j=\pm 3/2$. We therefore conclude, that the avoided crossing opened by the Rashba interaction
in the spectra of hexagonal flake that allowed for the fast spin flips induced by AC electric fields require intervalley scattering.


\section{Summary and Conclusion}
We have studied the possibility of the electrical control of the 
spin for a single excess electron confined within a silicene flake
using the  Rashba interaction. 
We found that the transitions within the lowest-energy quadruplet driven by an AC electric field -- also the spin-conserving ones -- require application of a strong vertical electric field  of the order of 1V/\AA. 
For $F_z=0$ the matrix elements for transitions with or without the spin flip vanish
due to cancellation of contributions of separate sublattices. The
field lifts the equivalence of the sublattices, the cancellation is no longer complete and the transitions involving both the spin and the orbital degrees of freedom within the quadruplet. 

The rate of the spin transitions is a nonmonotonic function of $F_z$ and becomes drastically increased within avoided crossing of the states of the quadruple with the states of the Kramers doublet. The spin flips occur then at the scale of several picoseconds in the external resonant AC field of a low amplitude of the order of hundreds volts per centimeter. However, due to the avoided crossing and the spin-conserving transitions within the same energy range, the induced spin flip is not a selective two-level Rabi transition, but the dynamics involves several eigenstates of the stationary Hamiltonian. 
 Generally,
outside the avoided crossings open by the Rashba interaction, the spin transitions occur at a slower rate but with the Rabi two-levels dynamics for the resonant frequency.
 We found that the properties of the flake as the source of the magnetic dipole moment can be controlled by the vertical electric field which quenches the currents circulating within the flake by localization of the wave function on a single sublattice.

The results were compared to the ones obtained for a circular quantum dot tailored within the flake by the spatial energy gap modulation. The confined states are separated from the edge
and the intervalley coupling is removed. Upon removal of the intervalley scattering the
avoided crossings that accelerated the spin flips are replaced by crossing of energy levels of opposite valleys and the speedup is no longer observed. Formation of the spin-valley doublets
in the absence of the intervalley scattering was demonstrated.

\section*{Acknowledgments}
This work was supported by the National Science Centre (NCN) according to decision DEC-2016/23/B/ST3/00821 and  by the Faculty of Physics and Applied Computer Science AGH UST statutory activities No. 11.11.220.01/2 within subsidy of the Ministry of Science and Higher Education, 
The calculations were performed on PL-Grid Infrastructure.


\begin{thebibliography}{00}
\bibitem{chow} S. Chowdhury and D. Jana, Rep. Prog. Phys. {\bf 79}, 126501 (2016).
\bibitem{Neto09}  A. H. Castro Neto, F. Guinea, N. M. R. Peres, K. S. Novoselov, and A. K. Geim, Rev. Mod. Phys. {\bf 81}, 109 (2009).
\bibitem{Zutic04} I. Zutic, J. Fabian, S. Das Sarma, Rev. Mod. Phys. {\bf 76}, 323 (2004).
\bibitem{Liu11} C.-C. Liu,  W. Feng, and Y. Yao, Phys. Rev. Lett. {\bf 107}, 076802 (2011).\bibitem{Liu} C.-C. Liu, J. Jiamg. and Y. Yao, Phys. Rev. B {\bf 84}, 195430 (2011).
\bibitem{Ezawa} M. Ezawa, Phys. Rev. Lett. {\bf 109}, 055502 (2012).
\bibitem{Pan14}  H. Pan, Z. Li, C.-C. Liu, G. Zhu, Z. Qiao, and Y Yao, Phys. Rev. Lett. {\bf 112}, 106802 (2014).
\bibitem{Xu12}   C. Xu, G. Luo, Q. Liu, J. Zheng, Z. Zhang, S. Nagase, Z. Gaoa,  and  J. Lu,  Nanoscale {\bf 4}, 3111 (2012).
\bibitem{Rachel14} S. Rachel and M. Ezawa, Phys. Rev B {\bf 89}, 195303 (2014).
\bibitem{Tsai13} W.-F. Tsai, C.-Y. Huand, T.-R. Chang, H. Lin, H.-T. Jeng, and A. Bansil, Nat. Comm. {\bf 4}, 1500 (2013).
\bibitem{nunez16}C. Nunez, F. Dominguez-Adame, P.A. Orellana, L. Rosales and R.A. Roemer, 2D Mater., {\bf 3}, 025006 (2016).
\bibitem{szak15} Kh. Shakouri, H. Simchi, M. Esmaeilzadeh, H. Mazidabadi, and F. M. Peeters, Phys. Rev. B {\bf 92}, 035413 (2015).
\bibitem{miso15} N. Missault, P. Vasilopoulos, V. Vargiamidis, F. M. Peeters, and B. Van Duppen, Phys. Rev. B {\bf 92}, 195423 (2015).
\bibitem{wu15} X.Q. Wu and H. Meng, J. Appl. Phys. {\bf 117}, 203903 (2015).
\bibitem{Ezawa12a} M. Ezawa, New J. Phys. {\bf 14}, 033003 (2012).
\bibitem{Tabert13}  C.J. Tabert and  E.J. Nicol, Phys. Rev. Lett. {\bf 110}, 197402 (2013).
\bibitem{romera} E. Romera and M. Calixto, EPL {\bf 111}, 37006 (2015).
\bibitem{Tao15} L. Tao, E. Cinqunta, D. Chappe, C. Grazianetti, M. Fanciulli, M. Dubey, A. Molle and D. Akinwande, Nat.  Nano. {\bf 10}, 227 (2015).
\bibitem{al2o3} M.X. Chen, Z. Zhong, and M. Weinert, Phys. Rev. B {\bf 94}, 075409 (2016).
\bibitem{nonmetal} M. Houssa, A. Stesmans, V.V. Afanasev, Interaction Between Silicene and Non-metallic Surfaces. In: Spencer M., Morishita T. (eds) Silicene. Springer Series in Materials Science, 
vol {\bf 235}, Springer, Cham (2016).
\bibitem{nonmetal1}
M. Houssa, G. Pourtois, V.V. Afanasev, A. Stesmans, Appl. Phys. Lett. {\bf 97}, 112106 (2010).
\bibitem{nonmetal2} Y. Ding, Y. Wang,  Appl. Phys. Lett. {\bf 103},
043114 (2013).
\bibitem{nonmetal3}  L.Y. Li, M.W. Zhao,  J. Phys. Chem.
C {\bf 118}, 19129 (2014)
\bibitem{Vogt12}  P. Vogt, P. De Padova, C. Quaresima, J. Avila, E. Frantzeskakis, M.C. Asensio, A. Resta, B. Ealet, and G. Le Lay, Phys. Rev. Lett. {\bf 108}, 155501 (2012).
\bibitem{me1}
A. Fleurence, R. Friedlein, T. Ozaki, H. Kawai, Y. Wang, Y. Takamura, Phys. Rev. Lett. {\bf 108}, 245501 (2012).
\bibitem{me2} C.C. Lee, A. Fleurence, Y. Yamada-Takamura, T. Ozaki, R. Friedlein, Phys. Rev. B {\bf 90}, 075422 (2014).
\bibitem{me3} L. Meng, Y. Wang, L. Zhang, S. Du, R. Wu, L. Li, Y. Zhang, G. Li, H. Zhou, W.A. Hofer, M.J.
Gao, Nano Lett. {\bf 13}, 685 (2013).
\bibitem{Aufray10} B. Aufray, A. Kara, S. Vizzini, H. Oughaddou, C. L\'eandri, B. Ealet, and G. Le Lay, Appl. Phys. Lett. {\bf 96}, 183102 (2010).
\bibitem{Feng12} B. Feng, Z. Ding, S. Meng, Y. Yao, X. He, P. Cheng, L. Chen, and K. Wu, Nano Lett. {\bf 12}, 3507 (2012).
\bibitem{Komputer} T.D. Ladd, F. Jelezko, R. Laflamme, Y. Nakamura, C. Monroe, and J.L. O Brien, Nature {\bf 464}, 45 (2010).
\bibitem{dl} D. Loss and D. P. DiVincenzo, Phys. Rev. A {\bf 57}, 120 (1998).
\bibitem{kl} C. Kloeffel and D. Loss, Annu. Rev. Condens. Matter Phys. {\bf 4}, 51 (2013).
\bibitem{meier} L. Meier, G. Salis, I. Shorubalko, E. Gini, S. Sch\"on, and K. Ensslin, Nat. Phys. {\bf 3}, 650 (2007).
\bibitem{edsr1} V. N. Golovach, M. Borhani, and D. Loss, Phys. Rev. B {\bf 74}, 165319 (2006).
\bibitem{edsr2} S. Debald and C. Emary, Phys. Rev. Lett. {\bf 94}, 226803 (2005).
\bibitem{edsr3} C. Flindt, A. S. Soerensen, and K. Flensberg, Phys. Rev. Lett. {\bf 97}, 240501 (2006).
\bibitem{edsr4} J. W. G. van den Berg, S. Nadj-Perge, V. S. Pribiag, S. R. Plissard, E. P. A. M. Bakkers, S. M. Frolov, and L. P. Kouwenhoven, 
Phys. Rev. Lett. {\bf 110}, 066806 (2013).
\bibitem{edsr5} K. C. Nowack, F. H. L. Koppens, Y. V. Nazarov, and
L. M. K. Vandersypen, Science {\bf 318}, 1430 (2007). 
\bibitem{edsr6} I. van Weperen, B. Tarasinski, D. Eeltink, V. S. Pribiag, S. R. Plissard, E. P. A. M. Bakkers, L. P. Kouwenhoven, and M. Wimmer
Phys. Rev. B {\bf 91}, 201413(R) (2015).
\bibitem{edsr7} F. Forster, M. M\"{u}hlbacher, D. Schuh, W. Wegscheider, and S. Ludwig, Phys. Rev. B {\bf 91}, 195417 (2015).
\bibitem{extreme} J. Stehlik, M. D. Schroer, M. Z. Maialle, M. H. Degani,
and J. R. Petta, Phys. Rev. Lett. {\bf 112}, 227601 (2014).
\bibitem{stroer} M. D. Schroer, K. D. Petersson, M. Jung, and J. R. Petta,
Phys. Rev. Lett. {\bf 107}, 176811 (2011).
\bibitem{edsrcnt1} F. Pei, E. A. Laird, G. A. Steele, and L. P. Kouwenhoven, Nature Nano. {\bf 7}, 630 (2012).
\bibitem{edsrcnt2} E. A. Laird, F. Pei, and L. P. Kouwenhoven, Nature Nano. {\bf 8}, 565 (2013).
\bibitem{edsrcnt3}T. Pei, A. Palyi, M. Mergenthaler, N. Ares, A. Mavalankar,J.H. Warner, G.A.D. Briggs, and E.A. Laird, Phys. Rev. Lett. {\bf 118}, 177701 (2017).
\bibitem{gru} M. Grujic, M. Zarenia, A. Chaves, M. Tadić, G. A. Farias, and F. M. Peeters
Phys. Rev. B {\bf 84}, 205441 (2011)
\bibitem{zarenia} M. Zarenia, A. Chaves, G. A. Farias, and F. M. Peeters, Phys. Rev. B {\bf 84}, 245403 (2011). 
 \bibitem{uwaga} note that the spin degree of freedom is not
considered in Ref. \cite{zarenia}.  
\bibitem{flake1} J. Fernandez-Rossier and J. J. Palacios, Phys. Rev. Lett. {\bf 99}, 177204 (2007).
\bibitem{flake2} B. Wunsch, T. Stauber, and F. Guinea, Phys. Rev. B  {\bf 77}, 035316 (2008).
\bibitem{flake3} A. D. Gucli, P. Potasz, and P. Hawrylak, Phys. Rev. B {\bf 82}, 155445 (2010).
\bibitem{flake4} P. Potasz, A. D. Guclu, and P. Hawrylak, Phys. Rev. B {\bf 81}, 033403 (2010).
\bibitem{flake5} W. L. Wang, O. V. Yazyev, S. Meng, and E. Kaxiras, Phys. Rev. Lett. {\bf 102}, 157201 (2009).
\bibitem{mori} S. Moriyama, Y. Morita, E. Watanabe, and D. Tsuya, Appl. Phys. Lett. {\bf 104}, 053108 (2014).  
\bibitem{buck0} K. Takeda and K. Shiraishi,  Phys. Rev. B {\bf 50}, 14916 (1994).
\bibitem{Cinquanta12} E. Cinquanta, E.  Scalise, D. Chiappe, C. Grazianetti, B. van den Broek, M. Houssa, M. Fanciulli, and Alessandro Molle,
 J. Phys. Chem. C {\bf 117}, 16719 (2013).
\bibitem{ni}Z. Ni, Q. Liu, K. Tang, J. Zheng, J. Zhou, R. Qin, Z.  Gao, D. Yu, and J. Lu,  Nano Lett. {\bf 12}, 113 (2012).
\bibitem{Drummond12} N.D. Drummond, V. Zolyomi, and V.I. Fal'ko, Phys. Rev. B 85, 075423 (2012).
\bibitem{kiku}  K. Kikutake, M. Ezawa, and Naoto Nagaosa, Edge states in silicene nanodisks, Phys. Rev. B {\bf 88}, 115432 (2013).
\bibitem{abdelsalam} H. Abdelsalam, M.H. Talaat, I. Lukyanchuk, M.E. Portnoi, and V. A. Saroka, J. Appl. Phys. {\bf 120}, 014304 (2016).
\bibitem{km} C.L. Kane and E.J. Mele, Phys. Rev. Lett. {\bf 95}, 226801 (2005).
\bibitem{revrev}K. Wakabayashi, Y. Takane, M. Yamamoto and M. Sigrist, New J. Phys. {\bf 11}, 095016 (2009).
\bibitem{km2} M. Laubach, J. Reuther, R. Thomale, and S. Rachel, Phys. Rev. B {\bf 90}, 165136 (2014).
\bibitem{szafran} B. Szafran, Phys. Rev. B {\bf 77}, 205313 (2008).
\bibitem{kuem} F. Kuemmeth, S. Ilani, D. C. Ralph, and P. L. McEuen, Nature
(London) {\bf 452}, 448 (2008).
\bibitem{waka} K. Wakabayashi, Phys. Rev. B {\bf 64}, 125428 (2001).
\bibitem{nori} G. Giavaras and F. Nori, Phys. Rev. B {\bf 83},  165427 (2011).
\bibitem{eosika} E. Osika and B. Szafran,
Phys. Rev. B {\bf 95}, 205305 (2017).

\end{thebibliography}
\end{document}